\documentclass[10pt]{article}
\usepackage{amsfonts,amssymb,latexsym,amsmath,amscd}
\usepackage[dvips]{graphics}
\usepackage{times,mathptm}
\usepackage{anysize}
\marginsize{1in}{1in}{1in}{1.3in}

\newcommand{\ket}[1]{|#1\rangle}
\newcommand{\M}[1]{\ket{\sf m#1}}

\title{An Arbitrary Two-qubit Computation In 23 Elementary Gates\thanks{
        Partially supported
        by the University of Michigan Mathematics department VIGRE summer
        stipend and the DARPA QuIST program.
        The views and conclusions contained herein are those of the authors
        and should not be interpreted as necessarily representing official
        policies of endorsements, either expressed or implied, 
        of respective funding institutions.
        }
       }
\author{Stephen S. Bullock and Igor L. Markov\\
        {\tt stephnsb@umich.edu \  imarkov@eecs.umich.edu}
}

\newenvironment{example}
    {
    \smallskip
    \refstepcounter{theorem}
    \noindent
    {\bf Example \arabic{section}.\arabic{theorem}} \ \ }
    {\hspace*{\fill}{$\Diamond$}
    \smallskip}

    {
    \smallskip
    \refstepcounter{theorem}
    \noindent
    {\bf Remark \arabic{section}.\arabic{theorem}} \ \ }
    {\hspace*{\fill}{$\Diamond$}
    \smallskip}

\newenvironment{definition}
    {
    \smallskip
    \refstepcounter{theorem}
    \noindent
    {\bf Definition \arabic{section}.\arabic{theorem}} \ \ }
    {\hspace*{\fill}{\ }
    \smallskip}

\newenvironment{proof}[1][]
    {
    \noindent
    {\bf Proof{#1}:  }
    }
    {\hspace*{\fill}{$\Box$}\smallskip}

    {
    \noindent
    {\bf Sketch{#1}:  }
    }
    {\hspace*{\fill}{$\Box$}\smallskip}

    {
    \noindent
    {\bf Reference:  }
    }
    {\hspace*{\fill}{$\odot$}\smallskip}

\newtheorem{theorem}{Theorem}[section]
\newtheorem{proposition}[theorem]{Proposition}

\newtheorem{corollary}[theorem]{Corollary}

\newsavebox{\Dgate}
\savebox{\Dgate}(0,0){
	\begin{picture}(0,0)
	{
        \put(0.5,0.5){\framebox(1,1){\large D}}
        \put(0,1){\line(1,0){0.5}}
        \put(1.5,1){\line(1,0){0.5}}
	}
	\end{picture}
}

\newsavebox{\Hadamardgate}
\savebox{\Hadamardgate}(0,0){
	\begin{picture}(0,0)
        \put(0.5,0.5){\framebox(1,1){\large H}}
        \put(0,1){\line(1,0){0.5}}
        \put(1.5,1){\line(1,0){0.5}}
	\end{picture}
}

\newsavebox{\Sgate}
\savebox{\Sgate}(0,0){
	\begin{picture}(0,0)
        \put(0.5,0.5){\framebox(1,1){\large S}}
        \put(0,1){\line(1,0){0.5}}
        \put(1.5,1){\line(1,0){0.5}}
	\end{picture}
}

\newsavebox{\UOneGate}
\savebox{\UOneGate}(0,0){
	\begin{picture}(0,0)
        \put(0.5,0.5){\framebox(1,1){\large $U_1$}}
        \put(0,1){\line(1,0){0.5}}
        \put(1.5,1){\line(1,0){0.5}}
	\end{picture}
}

\newsavebox{\UTwoGate}
\savebox{\UTwoGate}(0,0){
	\begin{picture}(0,0)
        \put(0.5,0.5){\framebox(1,1){\large $U_2$}}
        \put(0,1){\line(1,0){0.5}}
        \put(1.5,1){\line(1,0){0.5}}
	\end{picture}
}

\newsavebox{\UThreeGate}
\savebox{\UThreeGate}(0,0){
	\begin{picture}(0,0)
        \put(0.5,0.5){\framebox(1,1){\large $U_3$}}
        \put(0,1){\line(1,0){0.5}}
        \put(1.5,1){\line(1,0){0.5}}
	\end{picture}
}

\newsavebox{\UFourGate}
\savebox{\UFourGate}(0,0){
	\begin{picture}(0,0)
        \put(0.5,0.5){\framebox(1,1){\large $U_4$}}
        \put(0,1){\line(1,0){0.5}}
        \put(1.5,1){\line(1,0){0.5}}
	\end{picture}
}

\newsavebox{\UFiveGate}
\savebox{\UFiveGate}(0,0){
	\begin{picture}(0,0)
        \put(0.5,0.5){\framebox(1,1){\large $U_5$}}
        \put(0,1){\line(1,0){0.5}}
        \put(1.5,1){\line(1,0){0.5}}
	\end{picture}
}

\newsavebox{\USixGate}
\savebox{\USixGate}(0,0){
	\begin{picture}(0,0)
        \put(0.5,0.5){\framebox(1,1){\large $U_6$}}
        \put(0,1){\line(1,0){0.5}}
        \put(1.5,1){\line(1,0){0.5}}
	\end{picture}
}

\newsavebox{\Wgate}
\savebox{\Wgate}(0,0){
	\begin{picture}(0,0)
        \put(0.5,0.5){\framebox(1,1){\large W}}
        \put(0,1){\line(1,0){0.5}}
        \put(1.5,1){\line(1,0){0.5}}
	\end{picture}
}

\newsavebox{\Vgate}
\savebox{\Vgate}(0,0){
	\begin{picture}(0,0)
        \put(0.5,0.5){\framebox(1,1){\large V}}
        \put(0,1){\line(1,0){0.5}}
        \put(1.5,1){\line(1,0){0.5}}
	\end{picture}
}

\newsavebox{\Agate}
\savebox{\Agate}(0,0){
	\begin{picture}(0,0)
        \put(0.5,0.5){\framebox(1,1){\large A}}
        \put(0,1){\line(1,0){0.5}}
        \put(1.5,1){\line(1,0){0.5}}
	\end{picture}
}

\newsavebox{\Bgate}
\savebox{\Bgate}(0,0){
	\begin{picture}(0,0)
        \put(0.5,0.5){\framebox(1,1){\large B}}
        \put(0,1){\line(1,0){0.5}}
        \put(1.5,1){\line(1,0){0.5}}
	\end{picture}
}

\newsavebox{\Cgate}
\savebox{\Cgate}(0,0){
	\begin{picture}(0,0)
        \put(0.5,0.5){\framebox(1,1){\large C}}
        \put(0,1){\line(1,0){0.5}}
        \put(1.5,1){\line(1,0){0.5}}
	\end{picture}
}

\newsavebox{\diaggate}
\savebox{\diaggate}(0,0){
	\begin{picture}(0,0)
        \put(0.5,0.5){\framebox(1,1){ $R_{\bf z}$}}
        \put(0,1){\line(1,0){0.5}}
        \put(1.5,1){\line(1,0){0.5}}
	\end{picture}
}

\newsavebox{\TriGate}
\savebox{\TriGate}(0,0){
	\begin{picture}(0,0)
        \put(0.5,0.5){\framebox(1,1){\large 3}}
        \put(0,1){\line(1,0){0.5}}
        \put(1.5,1){\line(1,0){0.5}}
	\end{picture}
}

\newsavebox{\BiGate}
\savebox{\BiGate}(0,0){
	\begin{picture}(0,0)
        \put(0.5,0.5){\framebox(1,1){\large 2}}
        \put(0,1){\line(1,0){0.5}}
        \put(1.5,1){\line(1,0){0.5}}
	\end{picture}
}

\newsavebox{\BiGateBold}
\savebox{\BiGateBold}(0,0){
	\begin{picture}(0,0)
        \put(0.5,0.5){\framebox(1,1){\bf \large 2}}
        \put(0,1){\line(1,0){0.5}}
        \put(1.5,1){\line(1,0){0.5}}
	\end{picture}
}

\newsavebox{\UniGate}
\savebox{\UniGate}(0,0){
	\begin{picture}(0,0)
        \put(0.5,0.5){\framebox(1,1){\large 1}}
        \put(0,1){\line(1,0){0.5}}
        \put(1.5,1){\line(1,0){0.5}}
	\end{picture}
}

\newsavebox{\UniGateBold}
\savebox{\UniGateBold}(0,0){
	\begin{picture}(0,0)
        \put(0.5,0.5){\framebox(1,1){\bf \large 1}}
        \put(0,1){\line(1,0){0.5}}
        \put(1.5,1){\line(1,0){0.5}}
	\end{picture}
}

\newsavebox{\Xgate}
\savebox{\Xgate}(0,0){
	\begin{picture}(0,0)
        \put(0.5,0.5){\framebox(1,1){\large X}}
        \put(0,1){\line(1,0){0.5}}
        \put(1.5,1){\line(1,0){0.5}}
	\end{picture}
}

\newsavebox{\hwire}
\savebox{\hwire}(0,0){
	\begin{picture}(0,0)
	\put(0,1){\line(1,0){2}} 
	\end{picture}
}

\newsavebox{\botCNOT} 
\savebox{\botCNOT}(0,0){
	\begin{picture}(0,0)
        \put(1,3){\circle{1}}
        \put(0,1){\line(1,0){2}}
        \put(1,1){\line(0,1){2.5}}
    	\put(1,1){\circle*{0.4}}
	\put(0,3){\line(1,0){2}}
	\end{picture}
}

\newsavebox{\topCNOT} 
\savebox{\topCNOT}(0,0){
	\begin{picture}(0,0)
        \put(1,1){\circle{1}}
        \put(0,1){\line(1,0){2}}
        \put(1,0.5){\line(0,1){2.5}}
    	\put(1,3){\circle*{0.4}}
	\put(0,3){\line(1,0){2}}
	\end{picture}
}

\newsavebox{\topC}
\savebox{\topC}(0,0){
	\begin{picture}(0,0)
	\put(1,3){\circle*{0.4}}
	\put(1,3){\line(0,-1){1.5}}
	\put(0,3){\line(1,0){2}}
	\end{picture}
}

\newsavebox{\botC}
\savebox{\botC}(0,0){
	\begin{picture}(0,0)
	\put(1,1){\circle*{0.4}}
	\put(1,1){\line(0,1){1.5}}
	\put(0,1){\line(1,0){2}}
	\end{picture}
}

\begin{document}

\maketitle

\vspace{-4mm}
\begin{abstract}

 Quantum circuits currently constitute a dominant model for quantum
 computation \cite{NielsenC:00}.
 Our work addresses the problem of constructing quantum circuits to implement
 an arbitrary given quantum computation, in the special case of two qubits.
 We pursue circuits without ancilla qubits and as small a number of elementary 
 quantum gates \cite{BarencoEtAl:95,SongK:03} as possible. 
 Our lower bound for worst-case optimal two-qubit circuits calls for at least
 $17$ gates: $15$ one-qubit rotations and $2$ {\tt CNOT}s.
 To this end, we constructively prove
 a worst-case upper bound of $23$ elementary gates, of which at most 4
 ({\tt CNOT}s) entail multi-qubit interactions.  
 Our analysis shows that synthesis algorithms suggested in previous work,
 although more general, entail much larger quantum circuits than ours
 in the special case of two qubits. One such algorithm \cite{Cybenko:01}
 has a worst case of $61$ gates of which $18$ may be {\tt CNOT}s.
 
 Our techniques rely on the $KAK$ decomposition from Lie theory
 as well as the polar and spectral (symmetric Shur) matrix decompositions
 from numerical analysis and operator theory.
 They are related to the canonical decomposition of a two-qubit gate 
 with respect to the ``magic basis'' of phase-shifted Bell states
 \cite{KhanejaBG:01a,LewensteinEtAl:01}. We further extend this decomposition 
 in terms of elementary gates for quantum computation.

\end{abstract}
\vspace{-2mm}

\addtolength{\baselineskip}{-0.4pt}
\tableofcontents
\addtolength{\baselineskip}{0.4pt}

\section{Introduction}

  Quantum computations can be described by unitary matrices \cite{NielsenC:00}.
  In order to effect a quantum computation on a quantum computer,
  one must decompose such a matrix into a quantum circuit, which consists
  of elementary quantum gates \cite{BarencoEtAl:95} connected
  by Kronecker (tensor) and matrix products. Those connections are
  often represented using quantum circuit schematics. In some cases
  circuit decompositions require temporarily increasing the dimension
  of the underlying Hilbert space, which is represented by
  ``temporary storage lines''.
  Since there is always a multitude of valid circuit decompositions,
  one typically prefers those with fewer gates. 

  Algorithms for classical logic circuit synthesis \cite{HachtelS:00}
  read a Boolean function and output a circuit that implements the function
  using gates from a given gate library.
  By analogy, we can talk about quantum circuit synthesis. In this work
  we only discuss purely classical algorithms for such synthesis problems.
  Even at this early stage of quantum computing, it seems clear that
  algorithms for circuit synthesis are going to be as important
  in quantum computing as they are in classical Electronic Design Automation,
  where commercial circuit synthesis tools are necessary for the
  design of cellular phones, game consoles and networking chips.

  If a Boolean function is given by its truth table, then
  a two-level circuit, linear in the size of the truth table,
  can be constructed immediately.  Thus, it is the optimization
  of the circuit structure that makes classical circuit synthesis
  interesting. Given a unitary matrix, it is not nearly as easy
  to find a quantum circuit that implements it. Generic algorithms
  for this problem are known \cite{Tucci:99,Cybenko:01}, but in some
  cases produce very large circuits even when small circuits are possible.
  We hope that additional optimizations are possible.
  Importantly, the work in  \cite{Tucci:99} suggests that generic circuit
  decompositions can be found by means of solving a series of specialized
  synthesis problems, e.g., the synthesis of circuits consisting of NOT,
  CNOT and TOFFOLI gates as well as phase-shift circuits. Such specialized
  synthesis problems are addressed by other researchers
  \cite{BarencoEtAl:95,SongK:03,ShendePMH:02}.

  A recent work \cite{KhanejaBG:01a} on time-optimal control of
  spin systems presents a holistic view of circuit-related optimizations,
  which is based on the Lie group theory. However, their approach is not
  as detailed as previously published circuit synthesis algorithms,
  and comparisons in terms of gate counts are not straightforward.

  Our work can be compared to the GQC ``quantum compiler'' 
  \cite{GQC,BremnerEtAl:02a} available online.\footnote{
  We point out that the term ``compiler'' in classical computing
  means ``translator from a high-level description to a register-transfer
  level (RTL) description, e.g., machine codes''. The task of producing
  circuits with given function is commonly referred to as ``circuit
  synthesis''. In this context, digital circuits are called ``logic circuits''.}
  That program inputs a $4\times 4$ unitary $U$
  and returns a ``canonical decomposition'' which is not, in a strict
  sense, a circuit in terms of elementary gates. It also returns a circuit
  that computes CNOT using $U$ and one-qubit gates.
  When $U$ is used only once, this easily yields a circuit decomposition
  of $U$ in terms of elementary gates. However, it appears that
  not all input matrices can be processed successfully.\footnote{As of 
  March 2003, the quantum compiler \cite{GQC} fails on
$
  exp\left(
  i \left[
  \begin{array}{cccc}
     0& 1& 2& 3 \\
     1& 0& 4& 5 \\
     2& 4& 0& 6 \\
     3& 5& 6& 0 \\
  \end{array}
    \right]
  \right)
$. The authors are working on a bugfix and expect that
the problem lies in the code rather than the method.
}

  Our work pursues generic circuit decompositions
  \cite{BarencoEtAl:95,Cybenko:01}
  of two-qubit quantum computations up to global phase.
  While some authors consider arbitrary one-qubit gates elementary,
  we recall that they can be decomposed, up to phase, into a product
  of one-parametric rotations according to Equation \ref{eq:oneQbitDecomp}. 
  Therefore we only view the necessary one-parametric rotations as elementary. 
  Some of our results (constructive upper bounds) in terms of
  such elementary gates can be reformulated in terms of coarser elementary 
  gates. We also observe that the standard choice of elementary logic gates
  in classical computing ({\tt AND-OR-NOT}) was suggested in the $XIX^{th}$
  century by Boole for abstract reasons rather than based on specific
  technologies. Today the {\tt AND} gate is by far not the simplest 
  to implement in CMOS-based integrated circuits. This fact is addressed
  by commercial circuit synthesis tools by decoupling {\em library-less logic
  synthesis} from {\em technology-mapping} \cite{HachtelS:00}.
  The former uses an abstract gate library, such as {\tt AND-OR-NOT}
  and emphasizes the scalability of synthesis algorithms that capture
  the global structure of the given computation.
  The latter step maps logic circuits to a technology-specific gate library, 
  often supplied by a semiconductor manufacturer, and is based on local 
  optimizations. Technology-specific libraries may contain composite 
  multi-input gates with optimized layouts such as the {\tt AOI} gate
  ({\tt AND/OR/INVERTER}).

To this end, our algorithms are analogous to {\em library-less logic synthesis}.

{\bf Gate library.}
 We consider the following library of {\em elementary} one- and two-qubit gates:
\begin{itemize}
\item{ $R_y(\theta)=\left( \begin{array}{cc} \cos \theta/2 & \sin \theta/2 \\
-\sin \theta/2 & \cos \theta/2 \\ \end{array} \right)$  for all
$0 \leq \theta < 2\pi$;}
\item{ $R_z(\alpha)=\left( \begin{array}{cc} \mbox{e}^{-i\alpha/2} & 0 \\
0 & \mbox{e}^{i \alpha / 2} \\ \end{array} \right)$
 for all $0 \leq \alpha < 2 \pi$;}
\item{The CNOT gate, conditioned on either line.  }
\end{itemize}

  A given gate may, in principle, be applied to different lines.  We do not
restrict to which lines the above gates may be applied.  Note that
the gate library we use generates $U(4)$ up to global phase
\cite{Cybenko:01}.
  In order to find gate decompositions, we use the Lie-group techniques
  from \cite{KhanejaBG:01a}.
  The resulting procedure is often superior to previously published
  generic algorithms \cite{Tucci:99,Cybenko:01} in terms of the size
  of synthesized circuits.

\begin{theorem}
\label{thm:23}
Up to global phase, any two qubit computation may be realized
exactly by at most {\em twenty-three} elementary gates,
of which at most {\em four} are {\tt CNOT}s. No ancilla qubits are required.
\end{theorem}

   We do not know whether this result is optimal,
 but show that at least seventeen elementary gates are required.

   The remaining part of the paper is organized as follows.
 Section \ref{sec:background} covers the necessary background
 on quantum circuits and elementary gates for quantum computation
 \cite{BarencoEtAl:95}. Relevant matrix decompositions and prior work
 on circuit synthesis are described in Section \ref{sec:prior}, 
 including a related algorithm to decompose unitary
 matrices into elementary gates \cite{Cybenko:01}.  Section \ref{sec:entangler}
 introduces the ``magic basis'' from \cite{LewensteinEtAl:01},
 as well as the associated {\em entangler} and {\em disentangler} gates.
 In Section \ref{sec:23},
 we present a generic decomposition of an arbitrary two-qubit
 quantum computation into $23$ elementary gates or less using the $KAK$
 decomposition from Lie theory. We also give several examples.
 Lower bounds are discussed in Section \ref{sec:bounds},
 followed by conclusions and ongoing work in Section \ref{sec:conclusions}.

\section{Notation and Background}
\label{sec:background}
  $GL(2^k)=\{M\in (2^k\times 2^k)\mbox{-matrices}| \det(M)\neq 0 \}$.
  For $M\in GL(2^n)$, we consider its {\em adjoint} matrix $M^*$,
  produced from the transpose $M^t$ by conjugating each matrix element.
  $M$ is called {\em Hermitian} (synonym: {\em self-adjoint}) iff $M=M^*$.
  Hermitian matrices generalize symmetric real-valued matrices.

  Quantum states and quantum circuits are governed by the laws of quantum
  mechanics: $k$-qubit states are $2^k$-dimensional vectors, i.e., complex
  linear combinations of 0-1 bit-strings of length $k$.
  A quantum computation acting on $k$ qubits ($k$ inputs and $k$ outputs) 
  is modelled by a unitary $2^k\times 2^k$-matrix \cite{NielsenC:00}.
  We denote such matrices by
     $U(2^k)=\{M\in (2^k\times 2^k)\mbox{-matrices}|MM^*={\bf 1}\}$.
  $O(2^k)$ represents those matrices from $U(2^k)$ with real entries.
  $SU(2^k)$ and $SO(2^k)$ are the respective subsets with determinant one.
  Below, we will consider two generic elements of $SU(2)$:
  $A= \alpha E_{11} + (-\beta) E_{12} + \bar{\beta} E_{21} +
  \bar{\alpha}E_{22}$ and $B = \gamma E_{11} + (-\delta) E_{12} +
  \bar{\delta} E_{21} + \bar{\gamma}E_{22}$ with $1 =
  |\alpha|^2+|\beta|^2 = |\gamma|^2 + |\delta|^2$.
  Such a parameterization of $SU(2)$ can be verified directly.

  We largely ignore the effects of quantum measurement that is typically
  performed after a quantum circuit is applied, but we use
  the fact that any measurement is invariant under a global phase change.
  In mathematical terms, this means that any computation in
  $U(2^k)$ can be represented in normalized form by a matrix from $SU(2^k)$.

\subsection{Quantum circuits and elementary gates for quantum computation}

  In our work, we only discuss combinational quantum circuits, which
  are directed acyclic graphs where every vertex represents a gate.
  An output of a gate can be connected to exactly one input of another gate
  or one circuit output.  A similar restriction applies to gate inputs
  (see examples of quantum circuits in Figures \ref{fig:diag} and
  \ref{fig:entangler}).

  Following \cite{BarencoEtAl:95,SongK:03}, we attempt to express arbitrary
  computations using as small numbers of elementary gates as possible.
  In order to write matrix elements of particular gates,
  we order the elements of the computational basis lexicographically
  \cite{NielsenC:00}. The computation implemented by several gates
  acting independently on different qubits can be described by the
  Kronecker (tensor) product $\otimes$ of their matrices.
  In the usual computational basis $\ket{00}, \ket{01}, \ket{10},
\ket{11}$ ordered in the dictionary order, the matrix in $U(4)$
representing $A \otimes B$ (for $A$ and $B$ defined above) will
then be
\begin{equation}
\label{eq:normal}
(A \otimes B) =
\left(
\begin{array}{cc}
\alpha B & -\beta B \\
\bar{\beta} B & \bar{\alpha} B \\
\end{array}
\right)
\end{equation}

  Composition of multiple quantum computations is described 
  by the matrix product.
  However, as most circuit diagrams are read left-to-right,
  the order in respective matrix expressions is reversed. For example,
  the expression $(A \otimes B) (C \otimes D)$ corresponds to a two-qubit
  circuit where $C$ acts on the top line and $D$ on the bottom line,
  followed by $A$ acting on the top line and $B$ on the bottom line.
  Since the two lines do not interact, the same computation is performed by
  $AC$ acting on the top line and $BD$ acting on the bottom line
  independently, i.e., $(A \otimes B) (C \otimes D)= (AC \otimes BD)$.
  Sometimes this identity allows one to simplify quantum circuits and reduce 
  their gate counts.

  We distinguish two versions of the {\tt CNOT} gate,
 {\tt topCNOT} and {\tt botCNOT} conditioned on the top and bottom lines 
 respectively:
  (i) {\tt botCNOT} exchanges
      $\ket{01} \leftrightarrow \ket{11}$, i.e. {\tt CNOT} controlled
      by the top line, and
  (ii) {\tt topCNOT} exchanges $\ket{10} \leftrightarrow \ket{11}$.
  Those gates can be represented by matrices:
\begin{equation}
 \label{eq:CNOTs}
{\tt topCNOT} =
\left(
\begin{array}{cccc}
1 & 0 & 0 & 0 \\
0 & 1 & 0 & 0 \\
0 & 0 & 0 & 1 \\
0 & 0 & 1 & 0 \\
\end{array}
\right)
\quad \quad
{\tt botCNOT} =
\left(
\begin{array}{cccc}
1 & 0 & 0 & 0 \\
0 & 0 & 0 & 1 \\
0 & 0 & 1 & 0 \\
0 & 1 & 0 & 0 \\
\end{array}
\right)
\end{equation}

  An arbitrary one-qubit quantum computation
  can be implemented, up to phase, by three elementary gates.
  This is due to \cite[Lemma 4.1]{BarencoEtAl:95}, which
  decomposes an arbitrary $2\times 2$ unitary into
\begin{equation}
\label{eq:oneQbitDecomp}
U =
\left(
\begin{array}{cc}
\mbox{e}^{i\delta} &  0 \\ 0 & \mbox{e}^{i\delta} \\
\end{array}
\right)
\left(
\begin{array}{cc}
\mbox{e}^{-i\alpha/2} &  0 \\ 0 & \mbox{e}^{i\alpha/2} \\
\end{array}
\right)
\left(
\begin{array}{cc}
\cos \theta/2 &  \sin \theta/2  \\  - \sin \theta/2 & \cos \theta/2 \\
\end{array}
\right)
\left(
\begin{array}{cc}
\mbox{e}^{-i\beta/2} &  0 \\ 0 & \mbox{e}^{i\beta/2} \\
\end{array}
\right)
\end{equation}
To recover the non-$\delta$ parameters, we divide $U$ by its determinant.
The resulting matrix $\tilde{U}$ has $\delta = 0$, and
\begin{equation}
\tilde{U}^t
\left(
\begin{array}{cc}
0 & 1 \\
1 & 0 \\
\end{array}
\right)
\tilde{U} =
\left(
\begin{array}{cc}
- \mbox{e}^{-i \beta} \sin \theta & \cos \theta \\
\cos \theta & \mbox{e}^{i \beta} \sin \theta \\
\end{array}
\right)
\end{equation}
  We routinely ignore global phase because it does not
  affect the result of quantum measurement, which is the last step
  in quantum algorithms.
  A particular one-qubit computation, the Hadamard gate $H$,
  can be implemented, up to global phase, using two elementary gates
  as follows:
\begin{equation}
\label{eq:Hadamard}
H =
\frac{\sqrt{2}}{2}
\left(
\begin{array}{cc}
1 & 1 \\ 1 & -1 \\
\end{array}
\right)
=
\frac{\sqrt{2}}{2}
\left(
\begin{array}{cc}
-i & 0 \\ 0  & -i \\
\end{array}
\right)
\left(
\begin{array}{cc}
i & 0 \\ 0 & -i \\
\end{array}
\right)
\left(
\begin{array}{cc}
1 & 1 \\ -1 & 1 \\
\end{array}
\right)
\end{equation}
Similarly, the {\tt NOT} gate (also known as Pauli-$X$) requires
two elementary gates, up to a global phase:
\begin{equation}
\label{eq:NOT} 
X = {\tt NOT} = \left(
\begin{array}{cc}
0 & 1 \\
1 & 0 \\
\end{array}
\right)
=
\left(
\begin{array}{cc}
-i & 0 \\
0 & -i \\
\end{array}
\right)
\left(
\begin{array}{cc}
0 & 1 \\
-1 & 0 \\
\end{array}
\right)
\left(
\begin{array}{cc}
i & 0 \\
0 & -i \\
\end{array}
\right)
.
\end{equation}

\subsection{Circuits for diagonal unitaries}

For a diagonal matrix $D\in U(4)$, we have
$D = \mbox{diag}(z_1, z_2, z_3, z_4)$ with $z_i \bar{z}_i = 1, i = 1 \dots 4$.
The coordinates or their product can be normalized by choosing the global phase.
In contrast, the quantity $z_1 z_2^{-1} z_3^{-1} z_4$ is invariant.

\begin{proposition}
\label{prop:diagonaltrick}
i)  A diagonal matrix $D = \mbox{diag}(z_1, z_2, z_3, z_4)$
in $U(4)$ may be written as a tensor product of diagonal
elements of $U(2)$ iff  $z_1 z_2^{-1} z_3^{-1} z_4 = 1$.
ii)  Any gate which is diagonal when written in the computation basis may
be implemented up to phase in five elementary gates or less.
\end{proposition}

\begin{figure}
\begin{center}
\begin{picture}(8,4)
\put(0,0){\usebox{\hwire}}
\put(0,0){\usebox{\botCNOT}}
\put(2,2){\usebox{\Wgate}}
\put(2,0){\usebox{\hwire}}
\put(4,0){\usebox{\botCNOT}}
\put(4,0){\usebox{\hwire}}
\put(6,0){\usebox{\diaggate}}
\put(6,2){\usebox{\diaggate}}
\end{picture}

\caption{
\label{fig:diag}
Any $4\times 4$ diagonal unitary
$D = \mbox{diag}(z_1, z_2, z_3, z_4)$
may be decomposed into up to five elementary gates.
We set $\mbox{e}^{-i \phi} = z_1 z_2^{-1} z_3^{-1} z_4$ and
define $W = \mbox{diag}(\mbox{e}^{i \phi/4}, \mbox{e}^{-i\phi/4})$.
The two one-qubit unitaries on the right are diagonal.
Since the inverse of a diagonal matrix is also diagonal,
the form of this circuit can be reversed for any given matrix.
}
\end{center}
\end{figure}
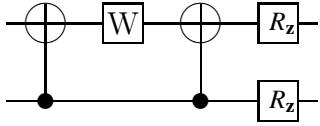

\begin{proof} i)  The forward implication follows from
$\mbox{diag}(\eta_1, \eta_2) \otimes
 \mbox{diag}(\eta_3, \eta_4) =
 \mbox{diag}(\eta_1 \eta_3, \eta_1 \eta_4, \eta_2 \eta_3, \eta_2 \eta_4)$.  \\
 For the reverse implication, rewrite that as
\[\mbox{diag}(\mbox{e}^{i \theta_1}, \mbox{e}^{i \theta_2}) \otimes
\mbox{diag}(\mbox{e}^{i \theta_3}, \mbox{e}^{i \theta_4}) =
\mbox{diag}(\mbox{e}^{i (\theta_1 + \theta_3)}, \mbox{e}^{i (\theta_1 + \theta_4)},
\mbox{e}^{i (\theta_2 + \theta_3)}, \mbox{e}^{i (\theta_2 + \theta_4)})\]
  If we are given the four diagonal entries $z_1 \ldots z_4$
  and wish to find $\theta$, this can be achieved by taking logarithms
  of $z_k$ and solving the resulting linear system
  in terms of $\theta_1, \dots, \theta_4$.
  The matrix of this $4\times 4$ system is degenerate and has rank 3.
  However, the constraint $z_1 z_2^{-1} z_3^{-1} z_4 = 1$ ensures
  that the system has a unique solution.

ii)  Consider the computation of Figure \ref{fig:diag}.
For a fixed $D = \mbox{diag}(z_1, z_2, z_3, z_4)$, put
$\mbox{e}^{-i\phi} = z_1 z_2^{-1} z_3^{-1} z_4$.  Now note the leftmost
three gates enact
\begin{equation}
\label{eq:diagconjbyCNOTs}
\left\{
\begin{array}{lcl}
\ket{00} \mapsto \mbox{e}^{i \phi/4} \ket{00} \\
\ket{01} \mapsto \mbox{e}^{-i \phi/4} \ket{01} \\
\ket{10} \mapsto \mbox{e}^{-i \phi/4} \ket{00} \\
\ket{11} \mapsto \mbox{e}^{i \phi/4} \ket{00} \\
\end{array}
\right.
\end{equation}
Thus by Equation \ref{eq:diagconjbyCNOTs} and part one of the present
proposition, the difference between $D$ and the leftmost
three gates is a pair of single elementary gates which
are diagonal elements of $U(1)\oplus U(1)$ on each line.
\end{proof}

\addtolength{\baselineskip}{2pt}

\section{Matrix Decompositions and Prior Work}
\label{sec:prior}

  As shown above, quantum circuits can be modelled by matrix formulas
  that decompose the overall computation (one large unitary matrix)
  into matrix products and tensor products of elementary gates (smaller
  unitary matrices). This suggests the use of matrix decomposition theorems
  from numerical analysis and Lie theory.
  Below, we revisit only decompositions relevant to our work:
 {\em SVD, polar, symmetric Shur (spectral), QR} \cite{GolubVL:96} 
  and {\em KAK} \cite{Knapp:98}.
  Additio\-nally, (i) a block-$2\times 2$ version of the {\em SVD}
  called the {\em CS} decomposition \cite[pp.77-79]{GolubVL:96}
  was used for circuit synthesis in \cite{Tucci:99},  and 
  (ii) the {\em LU} decomposition 
  \cite{GolubVL:96} was used to analyze CNOT-based circuits in \cite{BethR:01}.
   Most of those decompositions can be computed with existing 
   softare {\tt LAPACK}, downloadable from {\tt http://www.netlib.org}.

\subsection{Quantum circuit synthesis via the {\em QR} decomposition}

   The unitary matrix of a quantum computation can be analogized with
the truth table of a classical logic circuit. Logic minimization
aside, it is trivial to come up with a classical {\tt AND-OR-NOT}
circuit implementing a given truth table. Each line of the
truth table is implemented using {\tt AND} and {\tt NOT} gates,
then all lines are connected by {\tt OR} gates. The algorithm
proposed in \cite{Cybenko:01} solves a quantum version of this task.\footnote{
 We note that the work in \cite{Cybenko:01} to a large extent relies 
 on results in \cite{BarencoEtAl:95}.
}

   The algorithm relies on the theorem from numerical analysis, saying
that an arbitrary matrix can be decomposed into a product of a
unitary matrix {\tt Q} and an upper triangular matrix {\tt R},
not necessarily square \cite{GolubVL:96}. We are going to apply
this theorem to unitary matrices, which makes {\tt R} diagonal.
The canonical algorithm for {\tt QR}-decomposition is similar 
to the classical triangulation by row subtractions 
in that it zeroes out matrix elements one by one. 
Since elementary row operations are typically not unitary,
one instead applies a specially calculated element of $U(2)$ to a pair of rows
so as to zero out a particular matrix element. Such matrices are known as
{\em Givens rotations} and can be viewed as gates (not yet elementary) in a
quantum circuit for {\tt Q}.  This suggests that we find a decomposition for
the remaining diagonal component {\tt R}. Circuits for diagonal matrices are 
not explicitly addressed in \cite{Cybenko:01}, but are the subject of the work 
in \cite{HoggMPR:98}. The 2-qubit case addressed in the previous 
subsection is sufficient for further developments below.

Since each Givens rotation is a non-trivial two-qubit matrix, it should
be further decomposed into elementary gates. In the generic case,
the algorithm from \cite{Cybenko:01} entails one Givens rotation 
to nullify each matrix entry below the diagonal. 
Thus, a generic $4\times 4$ unitary representing a
$2$-qubit computation will decompose into six Givens rotations,
each uniquely determined. The first rotation ($G_{3,4}$ in \cite{Cybenko:01})
is between the states $\ket{10}$ and $\ket{11}$ whose indices corresponds 
to the last two rows of the matrix.  This rotation can be thought of as a
generic $1$-qubit rotation on the second qubit, controlled by the
first qubit. The work in \cite{BarencoEtAl:95,Cybenko:01} shows that such a
controlled rotation ${\tt topC}-V$ can be implemented using eight elementary
gates from the same gate library that we use. Namely, decompose $V$ according
to Equation \ref{eq:oneQbitDecomp} and use the parameters $\delta$,
$\alpha$, $\theta$ and $\beta$ to define matrices
\begin{equation}
  A=
\left(
\begin{array}{cc}
\mbox{e}^{-i\alpha/2} &  0 \\ 0 & \mbox{e}^{i\alpha/2} \\
\end{array}
\right)
\left(
\begin{array}{cc}
\cos (\theta/4) &  \sin (\theta/4)  \\
 - \sin(\theta/4) & \cos (\theta/4) 
\end{array}
\right)
\end{equation}

\begin{equation}
  B=
\left(
\begin{array}{cc}
\cos (-\theta/4) &  \sin (-\theta/4)  \\
 - \sin( -\theta/4) & \cos (-\theta/4) \\
\end{array}
\right)
\left(
\begin{array}{cc}
\mbox{e}^{i(\alpha+\beta)/4} &  0 \\ 0 & \mbox{e}^{-i(\alpha+\beta)/4} \\
\end{array}
\right)
\end{equation}
\begin{equation}
  C=
\left(
\begin{array}{cc}
\mbox{e}^{i(\alpha-\beta)/4} &  0 \\ 0 & \mbox{e}^{-i(\alpha-\beta)/4} \\
\end{array}
\right)
\end{equation}

\begin{equation}
  D=
\left(
\begin{array}{cc}
 1 &  0 \\ 0 & \mbox{e}^{i\delta} \\
\end{array}
\right)
\end{equation}

  One can verify that $ABC=I$ and $ATBTC=V/det(V)=\tilde{V}$.
  Therefore ${\tt topC}-V=(D\otimes {\bf 1}) \circ ({\bf 1} \otimes A)
  \circ {\tt topCNOT} \circ ({\bf 1} \otimes B) \circ {\tt topCNOT}
  \circ ({\bf 1} \otimes C)$. This decomposition is illustrated
  in Figure \ref{fig:cybenko} and implies 8 elementary gates because
  $A$ and $B$ require two each.

 The next Givens rotation  ($G_{2,3}$)
is between states $\ket{01}$ and $\ket{10}$.
It is not a controlled one-qubit rotation and thus more difficult 
to implement. The remaining Givens rotations are between $\ket{00}$
and $\ket{01}$ ($G_{1,2}$), $\ket{10}$ and $\ket{11}$ ($G_{3,4}$),
 $\ket{01}$ and $\ket{10}$ ($G_{2,3}$) as well as 
$\ket{10}$ and $\ket{11}$ ($G_{3,4}$). 
Four out of six are one-qubit rotations controlled by the top line ---
the most significant qubit.

To perform accurate gate counts in the 2-qubit case,
we first observe that an arbitrary 2-qubit diagonal matrix can
be implemented in five gates via Proposition \ref{prop:diagonaltrick}. 
Of those five two are {\tt CNOT}s.  The remaining effort is to count gates
in the six Givens rotations.
Following \cite{BarencoEtAl:95,Cybenko:01}, let $\mbox{\tt topC-}V$
be any $V \in U(2)$ controlled on the top line and acting on the second.
Then viewing a $4 \times 4$ matrix as block-$2 \times 2$, we obtain
\begin{equation}
\mbox{\tt topC-}V = \left(
\begin{array}{cc}
{\bf 1} & 0 \\
0 & V \\
\end{array}
\right) \mbox{ and } (X \otimes {\bf 1}) \circ \mbox{\tt topC-}V
\circ (X \otimes {\bf 1}) = \left(
\begin{array}{cc}
V & 0 \\
0 & {\bf 1} \\
\end{array}
\right)
\end{equation}

  Observe that {\tt topC-}$V$ implements $G_{3,4}$ and,
  according to Figure \ref{fig:cybenko}, costs eight gates,
  of which two are {\tt CNOT}s.
As shown in Equation \ref{eq:NOT}, inverters
cost two elementary gates each. Therefore the rotation $G_{1,2}$, 
implemented as above, costs twelve gates. 

With $\vec{0}$ being a two-column-high zero vector, the rotation $G_{2,3}$
can be implemented as
\begin{equation}
{\tt botCNOT} \circ {\tt topC-}(XVX) \circ {\tt botCNOT} =
\left(
\begin{array}{ccc}
1 & \vec{0}^t & 0 \\
\vec{0} & V & \vec{0} \\
0 & \vec{0}^t & 1 \\
\end{array}
\right) 
\end{equation}
  The computation ${\tt topC-}(XVX)$ considered as
  ${\tt topC-}\tilde{V}$ takes eight elementary gates, and
  thus $G_{2,3}$ can be implemented in ten elementary gates,
  of which four are {\tt CNOT}s.
 
  In the generic case, the algorithm from \cite{Cybenko:01} 
  is going to use {\em three} $G_{3,4}$ Givens rotations totalling 
  $24$ elementary gates of which $6$ are {\tt CNOT}s,
  {\em two} $G_{2,3}$ Givens rotations totalling $20$ elementary gates
  of which $8$ are {\tt CNOT}s and {\em one} $G_{1,2}$ Givens rotation
  which counts for $12$ elementary gates including $2$ {\tt CNOT}s.
  Additionally, we use $5$ elementary gates (of which $2$ are {\tt CNOT}s)
  to implement the diagonal {\tt R} via Proposition \ref{prop:diagonaltrick}.
  {\em Thus, $61$ gates will be required in the generic (worst) case,
   and $18$ of those will be {\tt CNOT}s}.

\subsection{Other matrix decompositions: {\em SVD}, polar, 
    symmetric Shur (spectral) and {\em KAK}}

  Golub and Van Loan \cite[p. 73]{GolubVL:96} define 
  the {\em Singular-Value Decomposition} (SVD) for complex matrices as follows:

\begin{definition}
  If $M\in \mathbb{C}^{m\times n}$, then there exist unitary matrices
  $U\in \mathbb{C}^{m\times m}$ and $V\in \mathbb{C}^{n\times n}$ such that
\[
  U^*MV=diag(\sigma_1,\ldots,\sigma_p)\in\mathbb{R}^{m\times n}
  \ \ \ p=\min\{m,n\}
\]
  where the $\sigma_i$ are singular values and
  $\sigma_1 \geq \sigma_2 \geq \ldots \geq \sigma_p \geq 0$.
  For real-valued $M$, $U$ and $V$ must be orthogonal.
\end{definition}
 
  In this work we are only interested in the case $m=n$,
  moreover, $n$ is typically a power of two.

\begin{definition}
  The {\em polar} decomposition of $M$ is $M=PZ$,
  where $Z$ is unitary and $P$ is Hermitian.
\end{definition}
  This can be derived from the {\em SVD} as follows \cite[p. 149]{GolubVL:96}. 
  If $M=U\Delta V^*$, then $M=(U\Delta U^*)(UV^*)=PZ$.  
  This decomposition is analogous to
  the factorization of complex numbers $z=\mbox{e}^{i\arg(z)}|z|$
  and intuitively similar to writing any complex $n\times n$ matrix
  as a sum of a Hermitian and skew-Hermitian matrices, in terms of 
  matrix elements: $m_{ij}=(m_{ij}+m_{ji}^*)/2+(m_{ij}-m_{ji}^*)/2$.
  Skew-Hermitian matrices exponentiate to unitaries, and Hermitian matrices 
  exponentiate to Hermitian. 
  However, in general $\exp(XY)\neq \exp(X)\exp(Y)$ unless $XY=YX$, 
  and polar decompositions cannot be computed by exponentiation.
  On the positive side, given an explicit $M$, $P^2$ can be computed as $M M^*$,
  and a possible $P$ can be found via matrix squareroot.
  In our work, we need a more refined version of the polar decomposition
  known from Lie theory \cite{Knapp:98}. The term {\em unitary polar}
  in the following definition is ours.

\begin{definition}
  The {\em unitary polar} decomposition of $M\in U(n)$ is
  $M=PZ$, where $Z\in SO(n)$ and $P=P^t$.
\end{definition}
 
\noindent 
  Since $Z$ and $M$ are unitary, so is $P$, demanding $P^{-1}=\bar{P}$.

\begin{definition}
  The {\em symmetric Shur} decomposition \cite[p. 393]{GolubVL:96},
  also known as the {\em spectral theorem} to operator theorists, 
  states that $M=O\Delta O^t$ where $M$ is a real-valued symmetric
  $n\times n$-matrix, $\Delta$ is diagonal and $O\in SO(n)$. 
  For a complex-valued Hermitian $M$, the matrix $O$ will have to be in $SU(n)$.
\end{definition}

  The symmetric Shur (spectral) decomposition can be interpreted as choosing 
  a basis in which $M$ is diagonal. Since such a basis must consist of 
  eigenvectors, the columns of $O$ list eigenvectors of $M$ in the initial 
  basis and $\Delta$ lists eigenvalues in the corresponding order.

\begin{proposition}
  \label{prop:superspectral}
    The following mild two-step generalization of the spectral theorem holds:
 \begin{enumerate}
  \item
   $\forall\ A, B$, symmetric real $n \times n$ matrices with
  $AB = BA$, $\exists \ O \in SO(n)$ such that $O A O^t$ 
  and $O B O^t$ are diagonal;
  \item $\forall \ P \in U(n)$ with $P=P^t$,
        $\exists\ O\in SO(n)$ such that $P=O\Delta O^t$, 
        where $\Delta$ is diagonal with norm-one entries.
 \end{enumerate}
\end{proposition}

\begin{proof}
 \begin{enumerate}
 \item
    It suffices to construct a basis which is simultaneously a basis of
eigenvectors for both $A$ and $B$.  Thus, say $V_\lambda$ is the $\lambda$
eigenspace of $B$.  For $v \in V_\lambda$, $B( Av) = A (Bv) = \lambda Av$, i.e.
$v \mapsto Av$ preserves the eigenspace.  Now find eigenvectors for $A$
restricted to $V_\lambda$, which remains symmetric.
 \item 
     Consider the real and imaginary parts of $P=A+iB$. Now 
$ {\bf 1} = P P^* = P \bar{P} = (A+iB)(A-iB) = (A^2 + B^2) + i(BA-AB)$.
Since the imaginary part of ${\bf 1}$ is ${\bf 0}$, we conclude that $AB=BA$.
The result follows from part 1.
 \end{enumerate}
\end{proof}

   The unitary polar decomposition and Proposition \ref{prop:superspectral}
   can be combined to produce the following variant of the $SVD$ for 
   unitary matrices. Suppose $U=PZ$ by the unitary polar decomposition.
   Apply Proposition \ref{prop:superspectral} to $P$ and write
   $U=PZ=O\Delta (O^t Z)=V\Delta W$ where $V,W \in O(n)$. Now multiply
   the first column of $V$ and the first entry of $\Delta$ by $\det(V)$,
   and then multiply the first row of $W$ and the first entry of 
   $\Delta$ by $\det(W)$. Thus we obtain $V,W \in SO(n)$.

\begin{definition}
  \label{def:KAK}
    The {\em normalized unitary KAK} decomposition 
    of $M\in U(n)$ is $M=V\Delta W$, where $V, W \in SO(n)$ 
    and $\Delta \in U(n)$ is diagonal. A related claim in terms of
    matrix groups is $U(n)=K A K$, where $K=O(n)$ and $A$ is the group of
    diagonal unitary matrices of determinant one.
\end{definition}

  The term {\em Lie theory}, in its modern use, refers to the mathematical
  theory of continuous matrix groups. Rather than study individual matrices, 
  as is common in numerical analysis, Lie theory studies collective behavior 
  of various types of matrices and often extends constructions
  from the group $GL(n)$ to its continuous subgroups such as $O(n)$ and $U(n)$.

  The {\em KAK} decomposition is a far-reaching generalization of the {\em SVD}
  and dates back to the origins of Lie theory in the 1920s.
  Knapp \cite[p.580]{Knapp:98} attributes it to Cartan \cite{Cartan:1927}.
  The $KAK$ decomposition of a {\em reductive} Lie group $G$ entails $G = KAK$
  where $K$ is a maximal proper {\em compact} subgroup and $A$ 
  is a {\em torus}. A {\em torus} is a connected Abelian group closed in $G$,
  and always a product of copies of the multiplicative group $(0,\infty)$ 
  and $U(1)$.  The $SVD$ decomposition can be seen as a special case with
  $G=GL(n,\mathbb{C})$,  $K=U(n)$ and the torus being the group of
  $n\times n$ diagonal matrices with positive real entries.
  In our work, we use another special case of the $KAK$ decomposition
  with $G=U(n)$,  $K=O(n)$ and the torus being the group of
  $n\times n$ diagonal unitary matrices. Section \ref{sec:magic}
  shows a surprising interpretation of $O(4)$ in terms of one-qubit gates.


\section{The Entangler Gate}
\label{sec:entangler}

 The {\em entangler} gate maps the computational basis into the ``magic basis'',
 which we introduce below. Together with its inverse ---
 the {\em disentangler} --- the entangler gate is useful for breaking down
 arbitrary two-qubit computations into elementary gates.  With such uses in
 mind, we implement the entangler and disentangler by elementary gates.

\subsection{$SU(2) \otimes SU(2) = SO(4)$ via the magic basis}
\label{sec:magic}

The ``magic basis'' \cite{LewensteinEtAl:01}
  provides an elegant way of thinking about
tensor products of one-qubit gates.\footnote{
Stated in terms of the Lie algebra of $U(4)$, this involves the isomorphism
$\mathfrak{u}(2) \oplus \mathfrak{u}(2) \cong \mathfrak{o}(4)$ \cite[p.
370]{Knapp:98}.
}

\begin{definition}
 The magic basis of phase shifted Bell states is given by
\begin{equation}
\label{def:magic}
\left\{
\begin{array}{lcl}
\M{1} & = & (\ket{00} + \ket{11})/\sqrt{2} \\
\M{2} & = & (i\ket{00} - i\ket{11})/\sqrt{2} \\
\M{3} & = & (i\ket{01} + i\ket{10})/\sqrt{2} \\
\M{4} & = & (\ket{01} - \ket{10})/\sqrt{2} \\
\end{array}
\right.
\end{equation}
Note that each is maximally entangled,
and the Arabic numbers are indices rather than energy states.
\end{definition}

Via a startling and omitted direct computation, the matrix
coefficients of $A \otimes B$ (in the notation of Equation
\ref{eq:normal}) with respect to the magic basis will all be real.
Hence $A \otimes B$ is orthogonal.  For example,
\begin{equation}
\label{eq:magicSample}
\begin{array}{lcl}
(A \otimes {\bf 1}) \M{1} & = & (\alpha \ket{00} + \bar{\beta} \ket{10} - \beta \ket{01}
  + \bar{\alpha} \ket{11})/\sqrt{2} \\
& = & \mbox{Re} \alpha \M{1} + \mbox{Im} \alpha \M{2} - \mbox{Im} \beta \M{3}
  - \mbox{Re} \beta \M{4} .\\
\end{array}
\end{equation}
Since changing basis does not change determinant, these
computations assert a ($U(4)$-inner) Lie-group isomorphism between
$SU(2) \otimes SU(2)$ and $SO(4)$. Importantly, both are known to
be connected \cite[p. 68]{Knapp:98}.

\begin{theorem}[from \cite{LewensteinEtAl:01}]
\label{thm:magic} The magic basis 
realizes the low dimensional isomorphism between $SU(2) \otimes SU(2)$ and
$SO(4)$. Specifically, for $V \in U(4)$ \emph{written with matrix
coefficients relative to the magic basis of Equation \ref{def:magic},}
\begin{equation}
[V \in SO(4) \subset U(4)] {\bf \Longleftrightarrow}
[(V:\mathbb{C}[\ket{0},\ket{1}] \otimes
\mathbb{C}[\ket{0},\ket{1}] \rightarrow \mathbb{C}) = A \otimes B \mbox{ for }
A, B \in SU(2)]
\end{equation}
\end{theorem}
Cf. Equation \ref{eq:normal} for the matrix for $A \otimes B$ in the
computational basis.

\begin{proof}
Continuing as in Equation \ref{eq:magicSample}, consider all $(A \otimes {\bf
1})\M{i}$ and $({\bf 1} \otimes B)\M{j}$ to show that $SU(2) \otimes SU(2)$
maps into $SO(4)$.  Now note $SU(2)$ is three dimensional since $|\alpha|^2 +
|\beta|^2 =1$, so $SU(2) \otimes SU(2)$ is six dimensional.  As $SO(4)$ has
$3+2+1$ real dimensions, this shows that the map defined above is onto the
identity component.
\end{proof}

\subsection{Definition and properties of $E$}

\begin{figure}
\begin{center}
\begin{picture}(12,4)
 \put(0,0){\usebox{\hwire}}
\put(0,0){\usebox{\botCNOT}}
\put(2,2){\usebox{\Sgate}}
\put(2,0){\usebox{\hwire}}
\put(4,0){\usebox{\botCNOT}}
\put(6,2){\usebox{\Hadamardgate}}
\put(6,0){\usebox{\hwire}}
\put(8,0){\usebox{\topCNOT}}
\put(10,0){\usebox{\botCNOT}}
\end{picture}

\caption{ 
\label{fig:entangler}
Implementing $E$ by elementary gates.
Here $S = \mbox{diag}(1,i)$ counts as one elementary gate
and the Hadamard gate $H$ counts as two.
}
\end{center}
\end{figure}
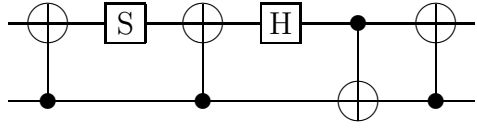

\begin{definition}
\label{def:entangler}
The {\em entangler} gate $E$ is the two qubit gate which maps the computational
basis into the magic basis:
$\ket{00} \mapsto \M{1}$, $\ket{01} \mapsto \M{2}$, $\ket{10} \mapsto \M{3}$,
and $\ket{11} \mapsto \M{4}$.
The inverse gate $E^*$ is called the {\em disentangler}.
\end{definition}

In terms of the computational basis, $E$ has the following matrix:
\begin{equation}
\label{eq:nTang}
 E =
\frac{\sqrt{2}}{2}
\left(
\begin{array}{cccc}
1 & i & 0 & 0 \\
0 & 0 & i & 1 \\
0 & 0 & i &-1 \\
1 & -i& 0 & 0 \\
\end{array}
\right)
\end{equation}
Now recall generally that for $g \in GL(n)$, a linear map $L$ with matrix $A$
subordinate to some basis $\{v_1, \dots v_n\}$ may also be expressed in terms
of the basis $\{g v_1, \dots g v_n\}$ via the conjugation map $A \mapsto g A
g^{-1}$.
In particular, $E$ is also given by the matrix above in the magic basis, 
and likewise $E^*$, and likewise any matrix commuting with $E$.  The typical
use of $E$ is the following Corollary of Theorem \ref{thm:magic}.

\begin{corollary}
\label{cor:tensor}
Suppose $V \in SO(4)$, that is $V \in U(4)$ with
$\mbox{det}(V)= 1$ and all real entries.  Then via the change of basis
remark of the last paragraph, $E V E^*$ is a tensor product of
one-line gates of the form of Equation \ref{eq:normal}.
\end{corollary}

One finds that $E$ can be realized up to global phase by seven
elementary gates, as shown in Figure \ref{fig:entangler}.
This is most easily verified by multiplying the
appropriate $4 \times 4$ matrices.  In particular,
Equation \ref{eq:CNOTs} writes {\tt topCNOT} and {\tt botCNOT}
as permutation matrices.  With that in mind, one can explicitly 
verify that
{ 
\begin{equation}
E =
  {\tt botCNOT} \circ
  {\tt topCNOT} \circ
\frac{\sqrt{2}}{2}
\left(
\begin{array}{cccc}
1 & 1 & 0 & 0 \\
1 & -1 & 0 & 0 \\
0 & 0 & 1 & 1 \\
0 & 0 & 1 & -1 \\
\end{array}
\right)
\circ {\tt botCNOT} \circ
\left(
\begin{array}{cccc}
1 & 0 & 0 & 0 \\
0 & 1 & 0 & 0 \\
0 & 0 & i & 0 \\
0 & 0 & 0 & i \\
\end{array}
\right)
  \circ {\tt botCNOT }
\end{equation}
}
Note that the circuit diagram in Figure \ref{fig:entangler}
travels right to left, so gate matrices are multiplied in reverse.
$S=\mbox{diag}(1,i)$ is an elementary gate up to global phase
$\mbox{e}^{- i \pi/4}$, and
the Hadamard gate $H$ can be implemented, up to global phase,
using two elementary gates as shown in Equation \ref{eq:Hadamard}.

In summary, $E$ requires four CNOT gates and three one-qubit rotations.
Similarly, $E^*$ may be implemented in seven elementary gates by writing
the inverse of each gate of Figure \ref{fig:entangler} in reverse order.

\addtolength{\baselineskip}{2pt}
\section{An Arbitrary Two-qubit Computation in $23$ Elementary Gates or Less}
\label{sec:23}

 In order to implement an arbitrary two-qubit computation with elementary
 gates, we first compute the normalized unitary $KAK$ decomposition 
 $U=K_1 A K_2$ of its unitary matrix $U$.
 According to the ``magic isomorphism'' from Section 
 \ref{sec:magic}, if we view $K_1$ and $K_2$ in the basis of Bell states,
 they decompose into tensor products of generic one-qubit computations,
 each requiring up to three one-qubit elementary gates. However, we then
 must view the remaining diagonal matrix in the same basis as well.
 The remaining part is of the form $E\Delta E^*$ for $\Delta$ diagonal and, 
 as shown below, can be implemented in $11$ gates due to its pattern of 
 zero entries.

\subsection{Decomposition algorithm}

The matrix decomposition implied by Theorem \ref{thm:23} is derived below,
and gate counts are in the next subsection.

\begin{proposition}
\label{prop:decomp}
Let $U$ be the matrix for any two-qubit computation in the computational basis,
so that $E U E^*$ represents $U$ in the magic basis.  Then
\begin{equation}
\label{eq:breakdown} U = (U_1 \otimes U_2) \circ {\tt botCNOT} \circ 
{\tt topC-}U_3
\circ 
({\bf 1} \otimes U_4 ) 
\circ 
{\tt botCNOT} \circ (U_5 \otimes U_6)
\end{equation}
where $U_1 \dots U_6$ are one-qubit gates on each line
and ${\tt topC-}U_3$ is controlled by the top line.
\end{proposition}

\begin{proof}
We are going to use the ``canonical decomposition'' of $U(4)$,
which is a combination of the $KAK$ decomposition of $U(4)$
and the ``magic isomorphism'' of Section \ref{sec:magic}.
The proof extends an algorithmic version of the canonical decomposition
towards elementary gates for quantum computation \cite{BarencoEtAl:95}
in the spirit of \cite{Cybenko:01}.

   In the algorithm below, steps 1-4 compute the normalized unitary $KAK$ 
  decomposition (see Definition \ref{def:KAK}) 
  of a given 2-qubit quantum computation $U$. Step 5 applies the magic
  isomorphism of Section \ref{sec:magic} to separate four generic
  one-qubit gates. Step 6 implements the remaining computation.

\begin{enumerate}
\item{
First, compute $P^2$ for $E^* U E = PK_1$ the unitary polar decomposition 
$P = P^t$, $K_1 \in SO(4)$.  To do so, note
$P^2 = P P^t = P K_1 K^t_1 P^t = E^* U E E^t U^t \bar{E}$.}
\item{
\label{item:eigenvectors}
  Now apply Proposition \ref{prop:superspectral} to $P^2$.
  This produces $P^2 = K_2 D K_2^{-1}$ for $K_2 \in O(4)$, $D$ diagonal.  
  Furthermore, choose $K_2 \in SO(4)$, so that $E K_2 E^*$ 
  is a tensor product via Corollary \ref{cor:tensor}.
}
\item{
\label{item:rootD}
Choose squareroots entrywise on the diagonal to form
$\sqrt{D}$, being careful to choose the signs of each root
 so that in the product $\mbox{det} \sqrt{D} = \mbox{det} U$.
This is in fact possible, since $\mbox{det}P^2 = (\mbox{det}U)^2$.
Having so chosen $\sqrt{D}$, compute $P = K_2 \sqrt{D} K_2^{-1}$.}
\item{One can now compute $K_1 = P^{-1} E^* U E = \bar{P} E^* U E$.  As
$\mbox{det} P = \mbox{det} U$, in fact $K_1 \in SO(4)$.}
\item{Thus $E^*UE = P K_1 = K_2 \sqrt{D} (K_2^{-1} K_1)$, whence
\[ U = (E K_2 E^*) (E \sqrt{D} E^* ) (E K_2^{-1} K_1 E^*) \]
upon conversion back to the computational basis.
Using Corollary \ref{cor:tensor}, we define $U_1, U_2, U_5$ and $U_6$ by
\begin{equation}
U_1 \otimes U_2 = E K_2 E^* 
\quad \textrm{ and } \quad
U_5 \otimes U_6 = E K_2^{-1} K_1 E^* 
\end{equation}
Both expressions may now be broken into explicit tensor products
of elements of $U(1)$. 
}
\item{What remains is to describe the implementation of $E \sqrt{D} E^*$.  
For this, label $\sqrt{D}= \mbox{diag}(a,b,c,d)$ with complex entries 
from $U(1)$.  Then
\begin{equation}
E \sqrt{D} E^* =
\frac{1}{2}
\left(
\begin{array}{cccc}
a+b & 0 & 0 & a-b \\
0 & c+d & c-d & 0 \\
0 & c-d & c+d & 0 \\
a-b & 0 & 0 & a+b \\
\end{array}
\right)
\end{equation}
Multiplying by a {\tt botCNOT} on the left flips rows two and four,
while multiplying on the right flips columns two and four. Thus,
\begin{equation}
E \sqrt{D} E^* = {\tt botCNOT} \circ
\left(
\begin{array}{cc}
U_4 & {\bf 0} \\
{\bf 0} & B \\
\end{array}
\right)
\circ 
{\tt botCNOT}
\end{equation}
for some $U_4, B \in U(2)$.  Choose $U_3$ so that $U_3 = B U_4^{-1}$.
Then the block-diagonal matrix $U_4 \oplus B$ may be implemented via
$U_4 \oplus B = 
({\bf 1} \oplus B U_4^{-1})
\circ
({\bf 1} \otimes U_4) 
= 
({\tt topC-}U_3)
\circ 
({\bf 1} \otimes U_4) 
$.}
\end{enumerate}
 Note that this algorithm has several unspecified degrees of freedom
 that may affect gate counts for specific 2-qubit computations.
 Arbitrary choices can be made in ordering eigenvectors 
 in step \ref{item:eigenvectors} and choosing a squareroot
 of a complex diagonal matrix in step \ref{item:rootD}. 
\end{proof}

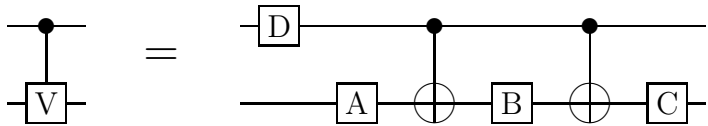
\begin{figure}
\begin{center}
\begin{picture}(18,2)
\put(0,0){\usebox{\topC}}
\put(0,0){\usebox{\Vgate}}
\put(3.5,2){\LARGE $=$}
\put(6,2){\usebox{\Dgate}}  
\put(6,0){\usebox{\hwire}}
\put(8,2){\usebox{\hwire}}   
\put(8,0){\usebox{\Agate}}
\put(10,0){\usebox{\topCNOT}}  
\put(10,2){\usebox{\hwire}}
\put(12,0){\usebox{\Bgate}}   
\put(12,2){\usebox{\hwire}}
\put(14,0){\usebox{\topCNOT}}  
\put(14,2){\usebox{\hwire}}
\put(16,0){\usebox{\Cgate}} 
\put(16,2){\usebox{\hwire}}
\end{picture}
\end{center}
\caption{ 
\label{fig:cybenko}
The implementation of a controlled-$V$ gate \cite[Figure 7]{Cybenko:01}.
The gates $A,B,C$ and $D$ are computed ibid. for a given $V$.
Here, $C$ and $D$  require one elementary gate each, 
while $A$ and $B$ require two each.
}
\end{figure}

\subsection{The overall gate decomposition and gate counts}

  Proposition \ref{prop:decomp} decomposes an arbitrary two-qubit unitary into
  $U = (U_1 \otimes U_2) \circ {\tt botCNOT}
\circ
     {\tt topC-}U_3
     \circ 
     ({\bf 1} \otimes U_4) 
\circ {\tt botCNOT} \circ (U_5 \otimes U_6)$
  where $U_1, \ldots, U_6$ are one-qubit gates.
  The immediate gate count yields: 
  \begin{itemize}
    \item three elementary rotations for each
          of five one-qubit gates $U_1, U_2, U_3, U_5$ and $U_6$, 
    \item two {\tt botCNOT} gates,
    \item eight elementary gates to implement the ${\tt topC-}U_4$ gate,
          according to \cite[Figure 7]{Cybenko:01}.
   \end{itemize} 
  The total gate count of 25 can be further reduced, given the structure 
  of the ${\tt topC-}V$ circuit in Figure \ref{fig:cybenko}.
  Indeed, that circuit can be written symbolically as
  ${\tt topC-}U_3 = ({\bf 1} \otimes C) \circ {\tt topCNOT} \circ 
  ({\bf 1} \otimes B) \circ {\tt topCNOT} \circ (D \otimes A) $.
  $C$ and $D$ are elementary gates up to phase,
  but $A$ and $B$ require up to two elementary gates \cite{Cybenko:01}.

  Since ${\tt topC-}U_3$ is next to $({\bf 1} \otimes U_4)$
  in Proposition \ref{prop:decomp},
  we can reduce $(D \otimes A) \circ ({\bf 1} \otimes U_4)$ to
  $( D \otimes U_7)$ where  $U_7=A U_4$. 
  By merging the computation $A$ with the generic one-qubit computation $U_4$
  that may require up to three elementary gates,
  one reduces the overall circuit by two elementary gates.

\begin{figure}
\begin{center}
\begin{picture}(18,4)
\put(0,2){\usebox{\TriGate}}  
\put(0,0){\usebox{\TriGate}}
\put(2,0){\usebox{\botCNOT}}  
\put(4,2){\usebox{\UniGate}}  
\put(4,0){\usebox{\TriGate}}
\put(6,0){\usebox{\topCNOT}}  
\put(8,0){\usebox{\BiGate}}   
\put(8,2){\usebox{\hwire}}
\put(10,0){\usebox{\topCNOT}} 
\put(12,0){\usebox{\UniGate}} 
\put(12,2){\usebox{\hwire}}
\put(14,0){\usebox{\botCNOT}} 
\put(16,2){\usebox{\TriGate}} 
\put(16,0){\usebox{\TriGate}}
\end{picture}

\caption{ 
\label{fig:all}
The decomposition of a generic 2-qubit quantum computation into 
up to $23$ gates.  Four generic one-qubit rotations are marked with ``3''
because they require up to three elementary gates. Computations requiring
two or one elementary gates are marked similarly.
}
\end{center}
\end{figure}
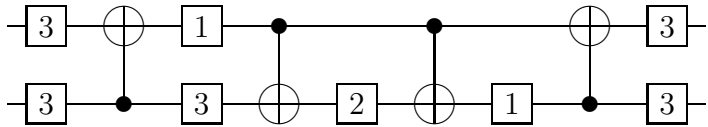

  The overall circuit decomposition can be described algebraically as follows:
\begin{equation}
   U =
 (U_1 \otimes U_2) \circ {\tt botCNOT} \circ 
      (D \otimes U_7)\circ
      {\tt topCNOT} \circ 
       ({\bf 1} \otimes B) \circ
       {\tt topCNOT} \circ 
      ({\bf 1} \otimes C) 
\circ 
{\tt botCNOT} \circ (U_5 \otimes U_6)
\end{equation}
 It is illustrated in Figure \ref{fig:all}, where gate counts are shown 
as well.

  Our circuit decomposition requires at most four CNOTs,
  while other gates are elementary one-qubit rotations. 
  Such a small number of non-one-qubit gates may be desired 
  in practical implementations where multi-qubit interactions
  are more difficult to implement.

  It is understood that Figure \ref{fig:all} and our gate counts
  refer to the worst case. Specific computations
  may require only some of those gates. In particular, the next
  section shows three examples that all require fewer gates
  than in the worst case. In those examples, our algorithm is able 
  to capture the structure of the given quantum computation.
  Unlike previously known circuit synthesis algorithms, ours
  can always implement $A\otimes B$ without using {\tt CNOT} gates.

\subsection{Examples} 
\label{sec:examples}

  Several examples below follow the algorithm from Theorem \ref{thm:23}.
  The order of eigenvectors and the choices of squareroots aimed at
  improving gate counts, but this search was not exhaustive.
  
\begin{example}  Let $H \otimes H$ be the two dimensional Hadamard
gate.  Following our algorithm, $E^* (H \otimes H) E \in SO(4)$, so that
$P^2 = {\bf 1}$ and we may choose $\sqrt{D}= P = {\bf 1}$ and
$K_2 = {\bf 1}$.  Then $K_1 = K_2^{-1} K_1 = E^* (H \otimes H) E$,
and the algorithm implements $H \otimes H$ in four elementary one-qubit
gates.  The {\tt CNOT}s cancel.  Similar comments apply to any $A \otimes B$. 
\end{example}

\begin{example}
Let $f:\mathbb{Z}/2\mathbb{Z} \rightarrow \mathbb{Z}/2\mathbb{Z}$
be the flip map, i.e. $f(n)=n+1$.  The Deutsch algorithm as described,
e.g., in \cite[p. 30]{NielsenC:00}, calls for a black-box gate $U_f$ 
with $U_f \ket{x} \ket{y} = \ket{x} \ket{y + f(x)}$, so that here
$U_f$ swaps $\ket{00}\leftrightarrow \ket{01}$.  Thus, $U_f$ is
easily implemented as $U_f = (X \otimes {\bf 1}) \circ {\tt topCNOT} \circ
(X \otimes {\bf 1})$ in five gates. Below, we decompose $U_f$ using our 
algorithm.

First, we find the Hermitian part 
of the unitary polar decomposition of $E^* U_f E$.
\begin{equation}
E^* U_f E E^t U_f^t \bar{E} = P P^t = P^2 =
\left(
\begin{array}{cccc}
0 & 0 & 0 & 1 \\
0 & 0 & 1 & 0 \\
0 & 1 & 0 & 0 \\
1 & 0 & 0 & 0 \\
\end{array}
\right)
\end{equation}
Now we must choose a basis of eigenvectors so as to diagonalize $P^2$.  Since
$P^2$ has both $\pm 1$ as double eigenvalues, there are uncountably many
ways to do this.  Simplifying things slightly, choose
\begin{equation}
K_2= \frac{\sqrt{2}}{2}
\left(
\begin{array}{cccc}
1 & 0 & 0 & 1 \\
0 & 1 & -1 & 0 \\
0 & 1 & 1 & 0 \\
-1 & 0 & 0 & 1 \\
\end{array}
\right)
\mbox{ so that } U_1 \otimes U_2 = E K_2 E^* = \frac{\sqrt{2}}{2}
\left(
\begin{array}{cc}
1 & -1 \\
1 & 1 \\
\end{array}
\right)
\otimes {\bf 1}
\end{equation}
Now the ordering of the column vectors of $K_2$ forces the diagonal
$D= \mbox{diag}(-1,1,-1,1)$ with $P^2 = K_2 D K_2^{-1}$.  We choose
$\sqrt{D} = (i,1,i,1)$, being careful to ensure $\mbox{det} \sqrt{D} =
\mbox{det} U_f = -1$.\footnote{Failing to do so will cause $\mbox{det}
K_1 \neq 1$ eventually, at which point $E K_2^{-1} K_1 E^*$ is not
a tensor product of one-qubit computations.}
Now putting $P = K_2 \sqrt{D} K_2^{-1}$, define $K_1 = \bar{P} E^* U_f E$.  
Then the one-line unitaries on the far side of the circuit may be computed as
\begin{equation}
E K_2^{-1} K_1 E^* =
\mbox{e}^{i \pi/4} \cdot \frac{1}{2}
\left(
\begin{array}{cccc}
i & 1 & 1 & -i \\
1 & i & -i & 1 \\
-i & -1 & 1 & -i \\
-1 & -i & -i & 1 \\
\end{array}
\right)
= \mbox{e}^{i\pi/4}
\frac{1}{\sqrt{2}}\left(\begin{array}{cc} i & 1 \\ -i & 1 \\ \end{array} \right)
\otimes \frac{1}{\sqrt{2}}
\left(\begin{array}{cc} 1 & -i \\ -i & 1 \\ \end{array} \right)
\end{equation}
To implement the latter in elementary matrices, one computes that
\begin{equation}
\left(\begin{array}{cc} i & 1 \\ -i & 1 \\ \end{array} \right)
=
\left(\begin{array}{cc} \mbox{e}^{i\pi/4} & 0 \\ 0 & \mbox{e}^{i\pi/4} \\ \end{array} \right)
\left(\begin{array}{cc} 1 & 1 \\ -1 & 1 \\ \end{array} \right)
\left(\begin{array}{cc} \mbox{e}^{i\pi/4} & 0 \\ 0 & \mbox{e}^{-i\pi/4} \\ \end{array} \right)
\end{equation}
while for the second factor similarly
\begin{equation}
\left(\begin{array}{cc} 1 & -i \\ -i & 1 \\ \end{array} \right)
=
\left(\begin{array}{cc} \mbox{e}^{i\pi/4} & 0 \\ 0 & \mbox{e}^{-i\pi/4} \\ \end{array} \right)
\left(\begin{array}{cc} 1 & 1 \\ -1 & 1 \\ \end{array} \right)
\left(\begin{array}{cc} \mbox{e}^{-i\pi/4} & 0 \\ 0 & \mbox{e}^{i\pi/4} \\ \end{array} \right)
\end{equation}

\begin{figure}
\begin{center}
\begin{picture}(20,4)
\put(0,0){\usebox{\hwire}}
\put(0,2){\usebox{\Xgate}}
\put(2,0){\usebox{\topCNOT}}
\put(4,2){\usebox{\Xgate}}
\put(4,0){\usebox{\hwire}}
\put(7.5,2){ \LARGE $=$}
\put(10,2){\usebox{\UFiveGate}}
\put(10,0){\usebox{\USixGate}}
\put(12,0){\usebox{\botCNOT}}
\put(14,0){\usebox{\UFourGate}}
\put(14,2){\usebox{\hwire}}
\put(16,0){\usebox{\botCNOT}}
\put(18,2){\usebox{\UOneGate}}
\put(18,0){\usebox{\hwire}}
\end{picture}
\end{center}

\caption{ 
\label{black}  
Diagrams for the $U_f$
black box for Deutsch's algorithm, where $f:\mathbb{Z}/2\mathbb{Z}
\rightarrow \mathbb{Z}/2\mathbb{Z}$ is $f(x)=x+1$.  The typical implementation
is shown at left, counting for $2+1+2=5$ gates and one {\tt CNOT}.  One
result of the current algorithm is shown at right.  Here,
$U_1 = R_y(-\pi)$, $U_4 = \mbox{e}^{-i\pi/4} R_z(-\pi/2) R_y(\pi) R_z(\pi/2)$,
$U_5 = i R_y(\pi) R_z (-\pi/2)$, and $U_6 = R_z(-\pi/2) R_y(\pi) R_z(\pi/2)$.
Thus this instance of the algorithm produces $10$ gates with two {\tt CNOT}s.
}
\end{figure}
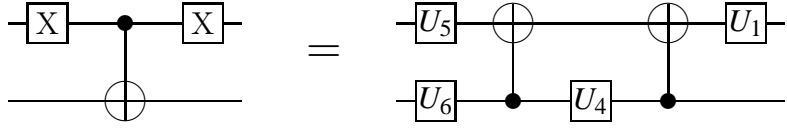

Finally, in this example the conditioned element is not required to 
implement $E \sqrt{D} E^*$.
Indeed,
\begin{equation}
{\tt botCNOT} \circ E \sqrt{D} E^*  \circ {\tt botCNOT} =
\mbox{e}^{-i \pi/4} \frac{\sqrt{2}}{2}
\left(
\begin{array}{cccc}
1 & i & 0 & 0 \\
i & 1 & 0 & 0 \\
0 & 0 & 1 & i \\
0 & 0 & i & 1 \\
\end{array}
\right)
= \mbox{e}^{-i \pi/4} {\bf 1} \otimes \frac{1}{\sqrt{2}}
\left(
\begin{array}{cc}
1 & i \\
i & 1 \\
\end{array}
\right)
\end{equation}
Thus, no conditioned gate is required within $E \sqrt{D} E^*$.  Moreover,
as we recently described the decomposition of the complex conjugate, we
see the ${\bf 1} \otimes U_4$ factor above counts for three gates.
Hence, our algorithm in this instance produces a decomposition with
$11$ rather than $5$ gates.  
It holds two {\tt CNOT}s rather than one {\tt CNOT}.
\end{example}

\begin{example}
One case of the algorithm also produces a $14$-gate decomposition of the
quantum Fourier transform $\mathcal{F}$, in contrast to the usual
$12$-gate implementation.  It has four rather than five two-qubit 
elementary gates ({\tt CNOT}s.)
Specifically, we write $\ket{00}, \dots \ket{11}$ as $\ket{0}, \dots, \ket{3}$.
Then the discrete Fourier transform $\mathcal{F}$ is given by
\begin{equation}
\label{eq:Fourier}
\ket{j}
{\buildrel \mathcal{F} \over \mapsto}
\frac{1}{2} \sum_{k=0}^3 (\sqrt{-1})^{jk} \ket{k}
\quad \quad
\mbox{ so that }
\quad \quad
\mathcal{F} = \frac{1}{2}
\left(
\begin{array}{cccc}
1 & 1 & 1 & 1 \\
1 & i & -1 & -i \\
1 & -1 & 1 & -1 \\
1 & -i & -1 & i \\
\end{array}
\right)
\end{equation}
Thus, the square of the Hermitian part of $E^* \mathcal{F} E$ is
\begin{equation}
E^* \mathcal{F} E E^t \mathcal{F}^t \bar{E} = P P^t = P^2 =
\left(
\begin{array}{cccc}
\mbox{e}^{i \pi/4} & 0 & 0 & \mbox{e}^{-i\pi/4} \\
0 & \mbox{e}^{i \pi/4} & \mbox{e}^{3 i \pi/4} & 0 \\
0 & \mbox{e}^{3 i \pi/4} & \mbox{e}^{i \pi/4} & 0 \\
\mbox{e}^{-i \pi/4} & 0 & 0 & \mbox{e}^{i \pi/4} \\
\end{array}
\right)
\end{equation}
Now we must diagonalize $P^2$.  As the eigenvalues are $1$ with
multiplicity two and $i$ with multiplicity two, there are infinitely
many possible eigen-bases of $\mathbb{C}^4$.  Choosing one such
for the columns of $K_2$ with determinant $1$, say
\begin{equation}
K_2= \frac{\sqrt{2}}{2}
\left(
\begin{array}{cccc}
1 & 0 & 0 & 1 \\
0 & 1 & -1 & 0 \\
0 & 1 & 1 & 0 \\
-1 & 0 & 0 & 1 \\
\end{array}
\right)
\mbox{ so that } U_1 \otimes U_2 = E K_2 E^* = \frac{\sqrt{2}}{2}
\left(
\begin{array}{cc}
1 & -1 \\
1 & 1 \\
\end{array}
\right)
\otimes {\bf 1}
\end{equation}
Now the ordering of the column vectors of $K_2$ forces the diagonal
$D= \mbox{diag}(i,i,1,-1)$ with $P^2 = K_2 D K_2^{-1}$.  The next
step is to choose $\sqrt{D}$ so that $\mbox{det} \sqrt{D} =
\mbox{det} E \mathcal{F} E^* = \mbox{det} \mathcal{F} = -i$.
Our choice is $\sqrt{D} = \mbox{diag}(\mbox{e}^{i\pi/4},
\mbox{e}^{i\pi/4},1,-1)$.  Then $P= K_2 \sqrt{D} K_2^{-1}$, so that
$K_1 = \bar{P} E^* \mathcal{F} E$ is complicated\footnote{Moreover, we had to
carefully choose $\mbox{det}\sqrt{D}=\mbox{det} \mathcal{F}$ to ensure
$\mbox{det} K_1 = 1$.  Otherwise $\mbox{det}K_2^{-1} K_1 \neq 1$
so that $E K_2^{-1} K_1 E^* \not\in U(2) \otimes U(2)$.}.  On the other hand,
\begin{equation}
K_2^{-1} K_1 =
\left(
\begin{array}{cccc}
\mbox{e}^{-i\pi/4} & 0 & \mbox{e}^{-i\pi/4} & 0 \\
0 & - \frac{\sqrt{2}}{2} & 0 & - \frac{\sqrt{2}}{2}\\
\frac{\sqrt{2}}{2} & 0 & - \frac{\sqrt{2}}{2} & 0 \\
0 & \mbox{e}^{-i\pi/4} & 0 & \mbox{e}^{i \pi/4} \\
\end{array}
\right)
\end{equation}
Thus, with some more matrix computations one computes that on the other side
\begin{equation}
U_5 \otimes U_6 =
E K_2^{-1}K_1 E^* =
[\mbox{diag}(1,\mbox{e}^{i\pi/4})
\circ H] \otimes \mbox{diag}(\mbox{e}^{-i\pi/4},-1)
\end{equation}
Note the first tensor factor would be more commonly referred to
as $T \circ H = \mbox{e}^{i\pi/8} R_z(\pi/4) (-i) R_z(-\pi) R_y(\pi)
= \mbox{e}^{-3\pi/8}R_z(\pi/4 - \pi) R_y(\pi)$.  On the other hand,
more commonly $U_6 = -T = (-1) \mbox{e}^{i\pi/8} R_z(\pi/4)$, so that
$U_5 \otimes U_6$ counts for $2+1=3$ gates.
This concludes the derivation of the outside one-line unitaries.

Finally, we implement $E \sqrt{D} E^*$.  The spacing of the zeroes
in $E \sqrt{D} E^*$ causes
${\tt botCNOT} \circ E \sqrt{D} E^* \circ {\tt botCNOT}$ to be block
diagonal, specifically (in $2 \times 2$ blocks)
\begin{equation}
\mbox{{\tt botCNOT}} \circ E \sqrt{D} E^* \circ \mbox{{\tt botCNOT}}=
\left(
\begin{array}{cc}
\mbox{diag}(\mbox{e}^{i \pi/4}, \mbox{e}^{i \pi/4}) & {\bf 0} \\
{\bf 0} & X \\
\end{array}
\right)
\end{equation}
Thus $U_4 = \mbox{diag}(\mbox{e}^{i \pi/4}, \mbox{e}^{i \pi/4})$
which is $={\bf 1}$ up to phase and does not cost any gates.
For $U_3$,
\begin{equation}
\mbox{e}^{-i \pi/4}X =
\left(
\begin{array}{cc}
\mbox{e}^{-3i\pi/4} & 0 \\
0 & \mbox{e}^{-3i\pi/4} \\
\end{array}
\right)
{\bf 1}
\left(
\begin{array}{cc}
0 & 1 \\
-1 & 0 \\
\end{array}
\right)
\left(
\begin{array}{cc}
-i & 0 \\
0 & i \\
\end{array}
\right)
\end{equation}
Thus, in the notation from the Section \ref{sec:background}
(and from \cite{BarencoEtAl:95,Cybenko:01}),
$\delta = -3 \pi/4$, $\alpha = 0$, $\theta = \pi$, and $\beta = \pi$.
Therefore the conditioned $\mbox{e}^{-3i\pi/4} X$ may be realized in 
$7$ gates.
As the unitaries $U_1, U_2={\bf 1}, U_5$ and $U_6$ (see Figure
\ref{fig:FT}(b)\ ) together require $5$ gates, we have 
$14$ gates total,
of which $2$ are {\tt botCNOT}s and $2$ are {\tt topCNOT}s.

\begin{figure}
\begin{center}
\begin{picture}(28,4)
\put(0,0){\usebox{\hwire}}
\put(0,2){\usebox{\Hadamardgate}}
\put(2,0){\usebox{\hwire}}
\put(2,2){\usebox{\Sgate}}
\put(2,0){\usebox{\botC}}
\put(4,0){\usebox{\Hadamardgate}}
\put(4,2){\usebox{\hwire}}
\put(6,0){\usebox{\topCNOT}}
\put(8,0){\usebox{\botCNOT}}
\put(10,0){\usebox{\topCNOT}}

\put(14.5,2){\LARGE $=$}

\put(18,2){\usebox{\UFiveGate}}
\put(18,0){\usebox{\USixGate}}
\put(20,0){\usebox{\botCNOT}}
\put(22,0){\usebox{\UThreeGate}}
\put(22,0){\usebox{\topC}}
\put(24,0){\usebox{\botCNOT}}
\put(26,2){\usebox{\UOneGate}}
\put(26,0){\usebox{\hwire}}
\end{picture}
\end{center}

\caption{ 
\label{fig:FT}
Shown are circuits for the Fourier transform: standard (left)
and produced by our algorithm (right). $U_1 = R_y(3 \pi)$,
$U_3 = \mbox{e}^{-i\pi/4}X$, $U_5 =
T H = \mbox{e}^{-3\pi/8}R_z(\pi/4 - \pi) R_y(\pi)$,
and $U_6 = -T = (-1) \mbox{e}^{i\pi/8} R_z(\pi/4)$.
Counting the conditioned $U_3$ as seven gates,
we get $2+1+1+7+1+2=14$ gates total.
}
\end{figure}
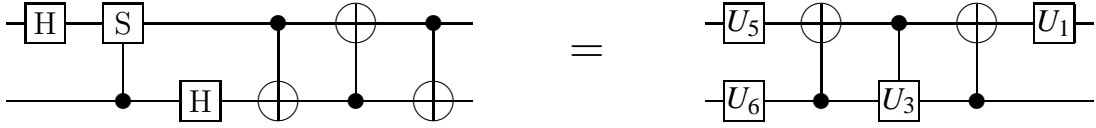

Compare the above 
to the standard $\mathcal{F}
= {{\tt botCNOT} \circ {\tt topCNOT} \circ {\tt botCNOT}} 
  ({\bf1} \otimes H) \circ ({\tt botC}-S) \circ (H \otimes {\bf 1})$
illustrated in Figure \ref{fig:FT}(a).
The conditioned $S$ can be implemented in $5$ gates 
as shown in Figure \ref{fig:cybenko}. Thus, the standard circuit
for the two-qubit Fourier transform has $12$ elementary gates.  
While this circuit has two gates fewer than the circuit produced
by our algorithm, it contains $5$ rather than $4$ {\tt CNOT} gates.
Since multi-qubit interactions are relatively expensive in many
quantum implementation technologies, the choice between the two
circuits may depend on specific technology parameters and 
implementation objectives.
\end{example}

\section{Gate Counts Versus Degrees of Freedom: Lower and Upper Bounds}
\label{sec:bounds}

   We have constructively shown in the previous section that any two-qubit 
  quantum computation can be implemented in 23 elementary gates or less,
  of which at most 4 are {\tt CNOT}s and remaining gates are one-qubit
  rotations.  As we do not know if this result can be improved,
  we show that at least $17$ elementary gates are required.

\begin{theorem}
   \label{th:17}
   There exists a two-qubit computation such that any circuit implementing it
   in terms of elementary gates consists of at least $17$ gates.
   In particular, $15$ one-qubit rotations are required and two {\tt CNOT}s.
\end{theorem}

\begin{proof}
   First, recall that two-qubit quantum computations
   can be represented by $4\times 4$-unitary matrices,
   and such matrices can be normalized to have determinant one
   because quantum measurement is not affected by global phase.
   Also recall that we use two types of elementary gates: (1) one-qubit 
   rotations with one real parameter each, and (2) {\tt CNOT}s which operate
   on two qubits and are fully specified (no parameters).

   Let us now consider the set $Q_C$ of quantum computations that can be 
   performed by some given two-qubit circuit $C$ with fixed topology, 
   where the parameters of one-qubit rotations are allowed to vary.  
   Fixed circuit topology means that [the graph of] connections between 
   elementary gates cannot be changed.  Since the overall unitary
   matrix can be expressed in terms of products and tensor products of
   the matrices of elementary gates, each matrix element is an infinitely
   differentiable function of the parameters of one-qubit rotations
   (more precisely, it is an algebraic function of $sin$ and $cos$
    of those parameters).  In other words, the set
   $Q_C$ is parameterized by one-qubit rotations
   and has the local structure of a differentiable manifold, whose topological
   dimension in $GL(4)$ is the number of one-qubit rotations in $C$
   with variable parameters. The topological dimension is  
   roughly-speaking the number of degrees of freedom.

   Since every computation can be implemented by a limited number of 
   elementary gates, the set of possible circuit topologies is finite.
   The set of all implementable quantum computations is a union 
   of sets $Q_C$ over the finite set of possible circuit topologies.
   Its topological dimension is the maximum of topological dimensions of 
   $Q_C$, i.e., the maximum number of one-qubit rotations with varying 
   parameters, allowed in one circuit.

   On the other hand, $\cup {Q_C}=SU(4)$. 
   We compute its topological dimension as follows.  First, we point out that
   the matrix logarithm (which is infinitely differentiable) maps $U(4)$ 
   one-to-one onto the set of skew-symmetric Hermitian matrices:
   $UU^*={\bf 1}\ \Rightarrow \log(U)+\log(U^*)=\log(U)+(\log(U))^*={\bf 0}$.
   Furthermore, $4\times 4$ skew-Hermitian matrices have $4$ independent reals
   on the diagonal and are otherwise completely determined by their $6$ complex
   upper-diagonal elements. Thus, the set of skew-Hermitian matrices has 
   topological dimension $16$, and the same is true about $U(4)$. 
   Subtracting $1$ for global phase, 
   we see that $15$ one-qubit rotations are needed to implement some two-qubit 
   computations. A randomly chosen computation is such with probability $1$,
   i.e., {\em almost always} rather than {\em always}.

   If no {\tt CNOT} gates are used in a given two-qubit circuit,
   the two lines never interact, and the two independent one-qubit
   computations can be implemented in $3$ elementary rotations each.
   Therefore, two-qubit computations implementable
   without {\tt CNOT}s have only $6$ degrees of freedom. Similarly,
   if only one {\tt CNOT} is allowed, then only $4\times 3=12$ rotations
   can be placed on two lines to the left and to the right of the {\tt CNOT} 
   to avoid gate reductions.
   This proves that at least $2$ {\tt CNOT} gates are necessary
   to implement any two-qubit computation requiring $15$ rotations.
\end{proof}
 
    Given the lower bound in Theorem \ref{th:17}, 
    the $19$ non-constant one-qubit rotations in Figure \ref{fig:all} 
    seem redundant as only $15$ rotations are required for dimension reasons.
    To this end, we offer another generic gate decomposition
    for arbitrary 2-qubit computations that entails no more than $15$
    non-constant one-qubit rotations, at the price of some constant
    rotations and significantly more {\tt CNOT} gates 
    than used by our main decomposition in Figure \ref{fig:all}.

  Recall from Proposition \ref{prop:decomp}
  that an arbitrary two-qubit unitary can be decomposed into 
  $U = (U_1 \otimes U_2) \circ (E D E^*) \circ (U_3 \otimes U_4)$
  where $U_1, \ldots, U_4$ are one-qubit gates and $D$ is a diagonal unitary.
  In this context, we use circuit decompositions for $E$, $E^*$ and $D$
  given in Sections \ref{sec:background} and \ref{sec:entangler}.
  The matrix $D$ is controlled by $3$ real parameters 
  ($4$ diagonal unitaries modulo global phase). It is implemented 
  in Figure \ref{fig:diag} using $3$ one-qubit rotations
  and $2$ {\tt CNOT}s. The entangler $E$ and disentangler $E^*$ are fixed
  matrices and require no parameters. The implementation of $E$ in 
  Figure \ref{fig:entangler} requires $3$ constant rotations 
  and $4$ {\tt CNOT}s.

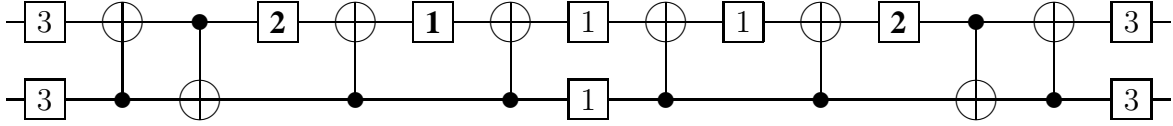
\begin{figure}
\begin{center}
\begin{picture}(34,4)
\put(0,2){\usebox{\TriGate}}
\put(0,0){\usebox{\TriGate}}
\put(2,0){\usebox{\botCNOT}}
\put(4,0){\usebox{\topCNOT}}
\put(6,2){\usebox{\BiGateBold}}
\put(6,0){\usebox{\hwire}}
\put(8,0){\usebox{\botCNOT}}
\put(10,2){\usebox{\UniGateBold}}
\put(10,0){\usebox{\hwire}}
\put(12,0){\usebox{\botCNOT}}
\put(14,2){\usebox{\UniGate}}
\put(14,0){\usebox{\UniGate}}
\put(16,0){\usebox{\botCNOT}}
\put(18,2){\usebox{\UniGate}}
\put(18,0){\usebox{\hwire}}
\put(20,0){\usebox{\botCNOT}}
\put(22,2){\usebox{\BiGateBold}}
\put(22,0){\usebox{\hwire}}
\put(24,0){\usebox{\topCNOT}}
\put(26,0){\usebox{\botCNOT}}
\put(28,2){\usebox{\TriGate}}
\put(28,0){\usebox{\TriGate}}
\end{picture}
\end{center}
  
\caption{ 
\label{fig:all28}
The overall structure entailed by our circuit decomposition.
Four generic one-qubit rotations are marked with ``3'' because 
they are worth up to three elementary gates. Two Hadamard gates 
are marked with ``2'' because they are worth two elementary gates.
Constant gates are in bold.
}
\end{figure}

  Adding up gate counts, we see that
  $U_1, \dots, U_4$ may require up to $12$ elementary gates alltogether.
  $D$ counts for $5$, while $E$ and $E^*$ count for $7$ each,
  for a total of $31$. However, upon inspection of the Figures \ref{fig:diag}
  and \ref{fig:entangler}, one notes that the circuit $E D E^*$ has two 
  canceling {\tt botCNOT} gates. Moreover, since the inverse of $D$ is, too,
  a diagonal unitary matrix, we can ``flip'' the asymmetric circuit for $D$
  in Figure \ref{fig:diag}. This allows us to merge a constant rotations 
  from $E$ with a variable rotation from $D$.
  The resulting circuit decomposition is illustrated in Figure \ref{fig:all28}
  and requires up to $28$ elementary gates total, of which $15$ are
  variable one-qubit rotations, $5$ are constant rotations and $8$ are
  {\tt CNOT}s. The slight asymmetry in  Figure \ref{fig:all28} is explained
  by the asymmetric circuit for $D$ in Figure \ref{fig:diag}.

  The following is a summary of our upper and lower bounds 
  for worst-case optimal 2-qubit circuits:
\begin{itemize}
   \item[(a)] an upper bound of $23$ elementary gates;
   \item[(b)] a lower bound of $17$  elementary gates.
   \item[(c)] an upper bound of $4$  {\tt CNOT} gates;
   \item[(d)] a lower bound of $2$   {\tt CNOT} gates;
   \item[(e)] an upper bound of $19$ one-qubit rotations;
   \item[(f)] an upper bound of $15$ variable elementary rotations;
   \item[(g)] a lower bound of $15$  variable elementary rotations;
\end{itemize}
  In our on-going work we show that three {\tt CNOT} gates are necessary
  and that the resulting lower bound of 18 elementary gates is tight.
  The implied decomposition contains at most 15 elementary rotations.

\section{Conclusions and On-going Work}
\label{sec:conclusions}

 It is a well-known result that any one-qubit computation
 can be implemented using {\em three} rotations or less \cite{BarencoEtAl:95}.
 Our work answers a similar question about arbitrary two-qubit computations
 assuming that {\tt CNOT} gates can be used in addition to single-qubit 
 rotations, without ancilla qubits. First, we show a lower bound that
 calls for at least {\em seventeen} elementary gates: {\em fifteen} rotations 
 and {\em two} {\tt CNOT}s.  We then constructively prove that
 {\em twenty three} elementary gates suffice to implement 
 an arbitrary two-qubit computation. At most {\em four} of those
  are {\tt CNOT}s and the rest are single-qubit gates. In comparison,
 a previously known construction \cite{BarencoEtAl:95,Cybenko:01} implies
 {\em sixty-one} gates of which {\em eighteen} are {\tt CNOT}s. While this
 construction is more general than ours, for two-qubit computations,
 our algorithm generates far fewer gates in the worst (generic) case.
 The savings in the number of multi-qubit gates ({\tt CNOT}s) are particularly
 dramatic. 

 In terms of techniques for the synthesis of quantum circuits,
 our work emphasizes the following general ideas:
\begin{itemize}
 \item changing the computational basis to maximally-entangled states
       by applying specially-designed gates
       with the purpose of recognizing quantum computations implementable 
       with one-qubit gates only;
 \item systematic use of matrix decompositions from numerical analysis and
       Lie theory: {\em polar}, {\em spectral} and {\em KAK};
 \item focus on matrix decompositions that are intrinsic to unitary matrices,
       e.g., {\em KAK} of $U(4)$, and include multiple non-trivial unitary 
       factors;
 \item incremental reduction of existing quantum circuits by local optimization;
       exploiting degrees of freedom in circuit synthesis may be useful to
       expose additional reductions.
\end{itemize}

  Specifically, we formalize the ``canonical decomposition'' of two-qubit
  computations \cite{LewensteinEtAl:01,KhanejaBG:01a} as an instance of 
  the {\em KAK} decomposition from Lie theory \cite{Knapp:98} for $U(4)$
  with $K=O(4)$ and $A$ diagonal. We propose an algorithm to compute 
  the {\em KAK} components and observe that elements of $O(4)$ can be 
  interpreted in the ``magic basis'' as pairs of one-qubit unitaries.
  Therefore, we change basis for all related matrices and further 
  decompose them into elementary gates for quantum computation.

  In our on-going work, with additional techniques,
  we are able to improve the lower bound to 18 elementary gates 
  and show that it is tight.

  We are also attempting to extend these ideas to three qubits or more.
  Two obstacles arise immediately:
\begin{itemize}
\item{Entanglement for three qubits is far more complicated than 
      it is for two qubits \cite{DuerEtAl:00}.
In particular, no known ``magic basis'' makes local unitaries tractable,
and there are distinct notions of maximally-entangled states.
}
\item{The use of the {\em KAK} decomposition does not automatically
      generalize beyond two qubits because 
$K \subset U(2^n)$ must be a sufficiently large subgroup,
in the sense that $U(2^n)/K$ must be a Riemannian symmetric space
\cite{Knapp:98,KhanejaBG:01a}.
Although both $O(4)$ and $U(2) \times U(2)$ are large subgroups of $U(4)$,
the set of local unitary gates $\otimes_{i=1}^{n} U(2)$
is not large enough in $U(2^n)$ for $n \geq 3$.
In particular, one \emph{does not} expect a decomposition of the type
$U_1 = U_2 D U_3$ for $U_1 \in U(8)$, $D$ diagonal, and $U_2$, $U_3 \in U(2)
\otimes U(2) \otimes U(2)$.} 
\end{itemize}

  With little hope for a direct matrix decomposition involving local unitaries,
  it remains possible, in principle, to construct a multi-step recursive 
  decomposition. A related example is available in \cite{Tucci:99}.


\addtolength{\baselineskip}{-2pt}
\section*{Acknowledgements}

  We thank Prof. Michael Nielsen (The University of Queensland)
  and Prof. Andreas Klappenecker (Texas A\&M University)
  for their feedback on earlier versions of this manuscript.


\begin{thebibliography}{99}
\bibitem{BarencoEtAl:95}
    A. Barenco et al.,
    ``Elementary Gates For Quantum Computation,''
    {\em Physical Review A} (52), 1995, 3457-3467.

 \bibitem{BethR:01} T. Beth and M. R\"{o}tteler,
 "Quantum Algorithms: Applicable Algebra and Quantum Physics,"
 {\em Springer Tracts in Modern Physics}, {\bf 173}, 2001, pp. 96-50.

\bibitem{BremnerEtAl:02a}
    M. J. Bremner, C. M. Dawson,  J. L. Dodd,
        A. Gilchrist, A. W. Harrow, D. Mortimer,
        M. A. Nielsen and T. J. Osborne,
    ``A Practical Scheme For Quantum Computation With Any Two-qubit
        Entangling Gate,'' {\tt quant-ph/0207072}, 2002.

\bibitem{Cartan:1927}
     \'{E}. Cartan, ``Sur Certaines Formes Riemanniennes
Remarquables des G\'{e}om\'{e}tries \`{a} Groupe Fondamental Simple,''
\emph{Annales Sci. \'{E}cole Norm. Sup.}, 44 (1927b), 345-467
(in {\OE}uvres Compl\`{e}tes, I, 867-989).

\bibitem{Cybenko:01}
    G. Cybenko, ``Reducing Quantum Computations to Elementary Unitary
    Operations,'' {\em Comp. in Sci. and Engin.}, March/April 2001,
    pp. 27-32.

\bibitem{DuerEtAl:00} W. D{\"u}r, G. Vidal and J.I. Cirac, 
 ``Three qubits can be entangled in two inequivalent ways,'' 
  {\em Physical Review A}, vol 62, 062314, 2000.

\bibitem{GQC}
      C. Dawson and A. Gilchrist,
      ``GQC: A Quantum Compiler'', 2002. \\
      {\tt http://www.physics.uq.edu.au/gqc/}

\bibitem{HachtelS:00}
    G. Hachtel and F. Somenzi, {\em Synthesis and Verification
    of Logic Circuits}, 3rd ed., Kluwer, 2000.

\bibitem{HoggMPR:98}
    T. Hogg, C. Mochon, W. Polak and E. Rieffel,
    ``Tools For Quantum Algorithms,''
     1998, {\tt quant-ph/9811073}.


\bibitem{GolubVL:96}
       G. H. Golub and C. F. Van Loan, {\em Matrix Computations},
       Johns Hopkins Press, Baltimore, 1996.

\bibitem{Knapp:98}
       A. W. Knapp, {\em Lie Groups Beyond an Introduction},
       Progress in Mathematics, vol. 140, {\em Birkh\"auser}, 1996.

\bibitem{KhanejaBG:01a}
       N. Khaneja, R. Brockett and S. J. Glaser,
       ``Time Optimal Control In Spin Systems,'' 2001
       {\tt quant-ph/0006114v2}.

\bibitem{LewensteinEtAl:01}
       M. Lewenstein, B. Kraus, P. Horodecki and I. Cirac,
     ``Characterization of Separable States and Entanglement
     Witnesses,'' {\em Physical Review A (3)}, vol. 63, no. 4, 2001,
     pp. 044304-7, {\tt quant-ph/0011050}.

\bibitem{NielsenC:00}
      M. Nielsen and I. Chuang,
      {\em Quantum Computation and Quantum Information},
      Cambridge Univ. Press, 2000.

\bibitem{SongK:03}
     G. Song and A. Klappenecker,
     ``Optimal Realizations of Controlled Unitary Gates,''
     2003. {\tt http://xxx.lanl.gov/abs/quant-ph/0301078}

\bibitem{Tucci:99}
     R. Tucci, ``A Rudimentary Quantum Compiler'', 1999,
     {\tt quant-ph/9902062}.

\bibitem{ShendePMH:02}
     V. V. Shende, A. K. Prasad, I. L. Markov and J. P. Hayes,
     ``Reversible Logic Synthesis'', to appear in
     {\em IEEE Trans. on Computer-Aided Design}, 2003,
     {\tt quant-ph/0207001}.

\end{thebibliography}

\end{document}